\pdfoutput=1

\documentclass[11pt]{article}

\usepackage{xcolor}
\usepackage{tabularx}
\usepackage{graphicx}
\usepackage{multicol}   
\usepackage{multirow}   
\usepackage{adjustbox} 
\usepackage{wrapfig} 
\usepackage{caption}   
\usepackage{subcaption}
\usepackage{makecell}
\usepackage{amssymb}
\usepackage{pifont}
\usepackage{xcolor}
\usepackage{colortbl}
\usepackage{enumitem}
\usepackage{booktabs}
\usepackage{algorithm} 
\usepackage{algpseudocode} 
\usepackage{amsmath} 
\usepackage{float}     
\usepackage[utf8]{inputenc}
\usepackage{multibib}
\usepackage{listings}

\usepackage{siunitx}       
\usepackage[table]{xcolor} 
\newcites{A}{References for Appendix}

\usepackage{tcolorbox}

\definecolor{LightGray}{HTML}{ececec}
\definecolor{LightBlue}{rgb}{0.8235,0.9737,0.9882}
\definecolor{LightGreen}{HTML}{effdee} 
\definecolor{LightYellow}{HTML}{fffee0} 
\definecolor{LightOrange}{HTML}{fee7da} 
\definecolor{textGreen}{HTML}{008000}
\definecolor{textRed}{HTML}{cc0000}
\definecolor{textBlue}{HTML}{0044FF} 
\definecolor{textSkyBlue}{HTML}{57EAD1} 

\usepackage[preprint]{acl}

\usepackage{times}
\usepackage{latexsym}

\usepackage[T1]{fontenc}

\usepackage[utf8]{inputenc}

\usepackage{microtype}

\usepackage{inconsolata}

\usepackage{graphicx}
\usepackage{url}
\usepackage{hyperref}
\usepackage{cleveref}
\usepackage[dvipsnames]{xcolor}
\title{Dynamic In-Group Persona Generation for Enhancing Human–AI Rapport}

\author{
Yoonseok Oh$^{1*}$, Inseo Jung$^{1*}$, Jinkyu Kim$^{1,2}$, Jungbeom Lee$^{1}$, Minwoo Kang$^{3\dagger}$, Suhong Moon$^{3\dagger}$\\
$^{1}$ Korea University\quad
$^{2}$ Kakao Mobility\quad
$^{3}$ University of California, Berkeley\\
\texttt{bd9983@korea.ac.kr} \\
{\footnotesize $^{*}$Equal contribution. \quad $^{\dagger}$Co-corresponding authors.}
}

\begin{document}
\maketitle

\begin{abstract}
LLM-based chatbots are increasingly applied in interpersonal domains such as counseling and peer support, where establishing human--AI rapport is crucial yet remains challenging. In this work, we introduce a novel approach for conditioning LLMs with \textbf{in-group personas}, which (i) first identifies a user’s primary concern and brief personal context (e.g., a computer science undergraduate worried about future career prospects), and (ii) generates a synthetic in-group persona that shares a similar primary concern while differing in background and narrative details, such as age or profession (e.g., a junior researcher at an AI startup).
Furthermore, we conduct a human-subject study to systematically 
evaluate the effectiveness of in-group persona agents in enhancing human–AI rapport. We compare our approach against two baseline conditions: a conventional agent without persona conditioning and an agent exhibiting minimal self-disclosure (e.g., ``I've felt that too''). 
Results from post-task questionnaires assessing rapport and user experience indicate that the in-group persona agent significantly improves perceived rapport and personal relevance compared to the baselines, and also yields more positive user experience—most notably higher engagement.
\end{abstract}
\section{Introduction}

\label{sec:intro}

Rapport, defined as ``mutual attention, positivity, and coordination''~\cite{tickle1990nature}, is crucial in counseling and peer support, where users must feel comfortable disclosing personal experiences~\cite{nienhuis2018therapeutic,truong2019role}. 
In supportive and interpersonal human--AI interactions such as peer-support or counseling-like conversations where users discuss personal concerns and may benefit from empathetic exchange, rapport can play a central role. Without a sense of relational connection, users may disengage, perceive the agent as impersonal, or hesitate to share meaningful information~\cite{liu2024chatgpt, pentina2023exploring}. In these contexts, rapport underpins not only conversational naturalness but also the effectiveness of AI systems in fostering user trust, cooperation, and sustained engagement.
\begin{figure}[t]
  \centering
\includegraphics[width=\linewidth]{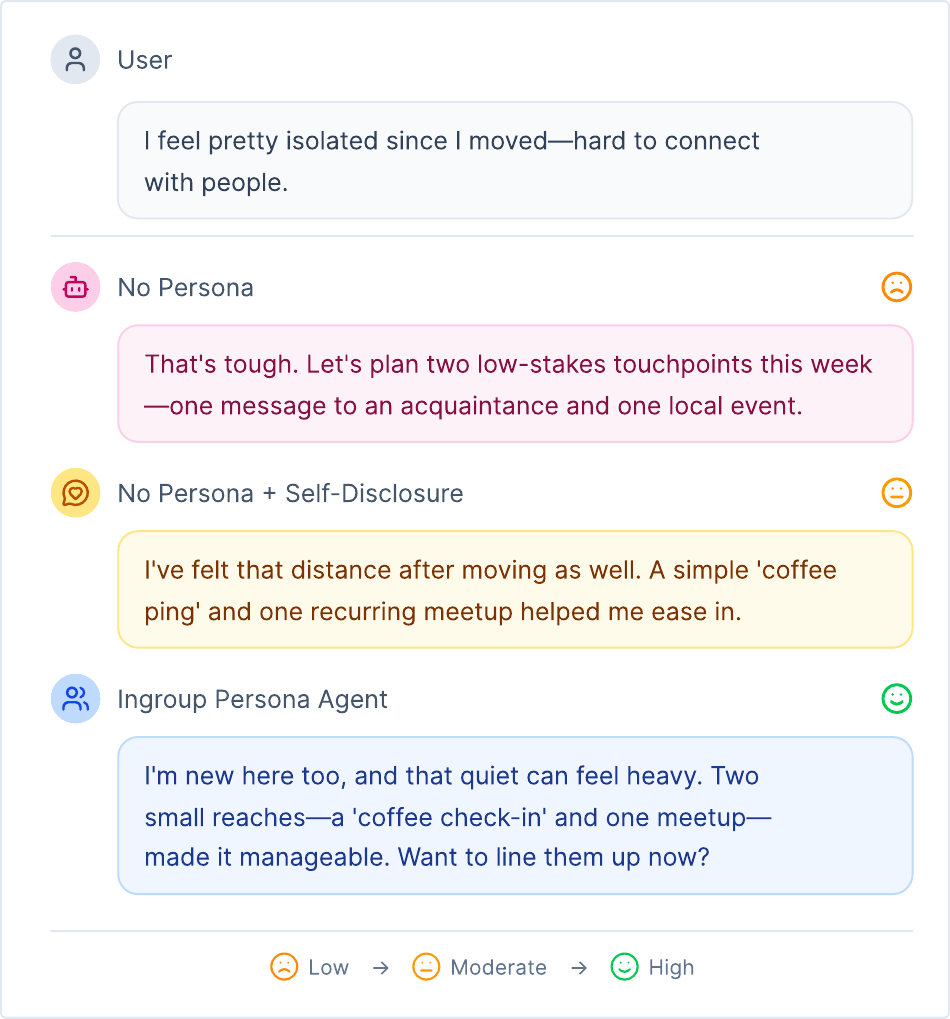}
  \caption{
  \textbf{Dynamic Persona Generation for Rapport in Human-AI Dialogue.} Comparison of three LLM agents responding to the same user utterance expressing life concern.
  \emph{No Persona}, the chatbot LLM without additional persona conditioning, gives neutral, generic guidance; \emph{No Persona + Self-Disclosure} adds a brief, generic self-disclosure that is not tailored to the user's elicited context; finally,
  \emph{In-group Persona Agent (Ours)} uses a concern-aligned persona derived from pre-chat context to provide contextualized self-disclosure. 
  }
  \label{fig:teaser}
\end{figure}

While large language model (LLM)-based chatbots are increasingly deployed in domains such as mental health support, education, and customer service, establishing \textit{rapport} with users remains a central challenge.
Despite that we are witnessing rapid deployment of AI-based counseling and mental health support systems, such as Woebot and Wysa~\cite{fitzpatrick2017delivering, inkster2018empathy}, and test trials in clinical contexts ~\cite{heinz2025randomized, macneill2024effectiveness}, chatbots often fail to demonstrate contextual understanding of social cues required to form rapport comparable to human counselors~\cite{sands2021managing, chan2022challenges, klein2025effects}.
As a result, LLM-based systems still face limitations in maintaining long-term effectiveness: sustained success appears contingent on users perceiving the agent as trustworthy, empathetic, and engaging~\cite{siddals2024happened, limpanopparat2024user, liu2025illusion}. 

A key mechanism for fostering rapport is \textit{self-disclosure}, which refers to the act of sharing personal thoughts, feelings, or experiences to build trust and intimacy. Interpersonal research shows that disclosure strengthens trust and intimacy via reciprocity~\cite{carpenter2015social}. In Human-Computer Interaction (HCI), chatbots that share personal or emotional information elicit greater user self-disclosure and satisfaction~\cite{ho2018psychological, lee2020hear}, and emotional self-disclosure has been shown to increase reuse intention~\cite{park2023effect}. 

Another mechanism to build rapport is to show \textit{similarity and in-group preference}: according to similarity--attraction theory~\cite{byrne1972attraction}, people respond more positively to partners who resemble them. This effect can be generalized to chatbots, where perceived similarity in personality or style enhances trust and affinity~\cite{reeves1996media, jin2023birds}. In-group categorization further amplifies these effects, with users treating agents as more trustworthy and cooperative when framed as group members~\cite{nass1997machines, eyssel2012social}. Yet, most chatbot studies have emphasized \textit{shallow similarity} (e.g., tone or trait matching) rather than shared lived experiences or social identity, leaving the deeper role of \textit{contextual alignment} underexplored~\cite{ahn2021ai, alawi2023accepting}.  

In this work, we propose a framework of \emph{dynamic persona generation} that grounds LLM personas on users’ expressed concerns and contexts, forming the basis of our \textbf{In-group Persona Agent (IPA)}.
This framework aims to improve relational outcomes in human--AI interaction through careful agent persona construction and concern-grounded alignment.
Our approach operationalizes findings from peer support research in mental health highlighting how shared experiences between interlocutors provide unique benefits, hope, validation, and a sense of not being alone~\cite{repper2011review}. 
We model users’ expressed \textit{concerns}--issues they currently care about that capture their emotional and goal-oriented states~\cite{randle2017societal,ho2018psychological,cortland2017solidarity}--that are in turn utilized to form user-centered experiential representations that inform the agent’s persona design.

With user concern-based experiences incorporated as part of the persona, our LLM agents are shown to produce responses that are both empathetic and contextually appropriate.
To evaluate this framework, we conducted interactive sessions simulating everyday conversations where participants discussed personal concerns such as career and employment issues in a supportive but non-professional context. 
These dialogues provided a balanced setting where users sought both emotional empathy and practical perspectives, allowing us to assess how concern-aligned personas influence relational outcomes in human-AI interaction.


\paragraph{Contributions.} We present the following contributions:

\begin{itemize}
\vspace{-0.5em}
    \item We propose \textbf{In-group Persona Agent (IPA)}, a method that dynamically integrates persona generation and conversational inference through a multi-stage prompt pipeline, constructing concern-aligned personas from elicited user dialogue context.
    \vspace{-0.5em}

    \item We complement the human-subject evaluation with post-hoc analyses that validate persona quality and characterize turn-level and conversation-level interaction patterns, thereby shedding light on the behavioral mechanisms underlying rapport gains.
    \vspace{-0.5em}


    \item Controlled evaluations show that IPA improves rapport (especially personal relevance) over baselines, with modest UX gains most evident in engagement, highlighting \textit{in-group alignment} as a key factor in human--AI relational quality.
\end{itemize}



\begin{figure*}[!t]
    \centering
    \includegraphics[width=0.85\linewidth]{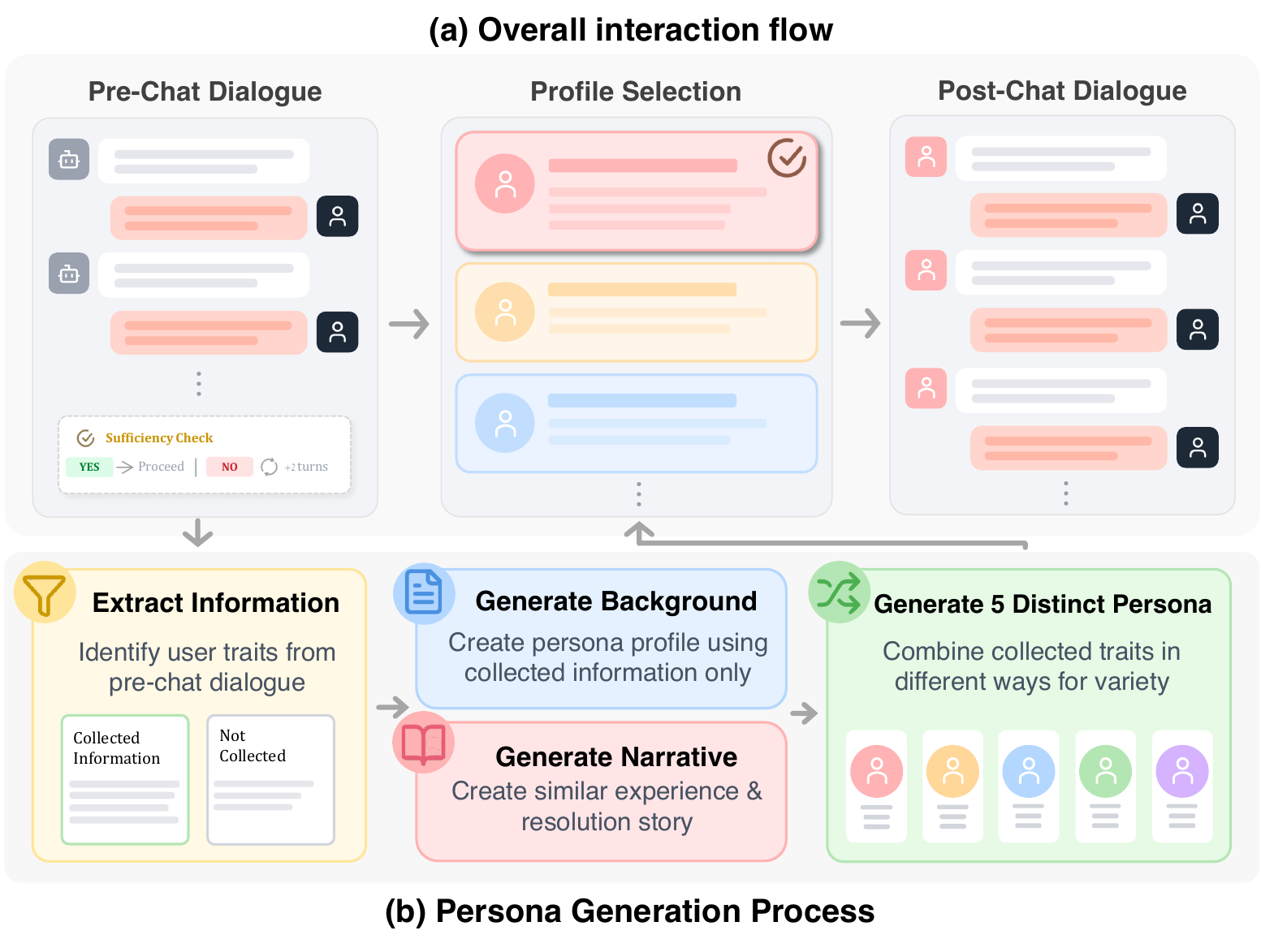}
      \caption{\textbf{Overview of the proposed \textit{In-Group Persona Agent}.} 
          \textbf{(a) Overall interaction flow:} a pre-chat dialogue elicitation with a sufficiency check collects the user’s concern; the user then selects one of five candidate personas; the main conversation proceeds with the selected persona 
          (see \Cref{sec:method}). 
          \textbf{(b) Persona generation process:} traits are extracted from the pre-chat dialogue and partitioned into \emph{collected} vs.\ \emph{not collected}; using both information, the system composes a background and a narrative (similar experience and resolution) and combines the traits in different ways to produce five distinct personas.}
    \label{fig:main_arch}
\end{figure*}
\section{Related Work}
\label{sec:relwork}
\subsection{LLM Persona Conditioning}
Large Language Models (LLMs) can be steered via prompt-only conditioning to display persona-like linguistic patterns without parameter updates. Evidence to date is strongest within-session and is task-dependent, indicating reliable short-horizon style control rather than persistent identity \cite{moon2024virtual,kang2025deep,jiang2023personallm,tseng2024two}. Backstory-style prompts improve response consistency and distributional alignment in large-scale simulations \cite{moon2024virtual,kang2025deep}, while personality prompts (e.g., Big Five) induce trait-congruent tendencies across tasks under matched instructions \cite{jiang2023personallm}. Recent evaluation work further studies LLM agents conditioned on assigned personas, assessing whether their actions, linguistic habits, and responses remain consistent with the given persona across diverse environments \cite{samuel2024personagym}.

Unlike previous studies that relied on static backstories or trait-based prompts to elicit transient stylistic alignment, our In-group Persona Agent dynamically generates concern-aligned personas grounded in users’ expressed contexts, enabling sustained relational alignment and measurable improvements in rapport and engagement.

\subsection{Self-Disclosure by AI Chatbots}
Research on interpersonal dynamics and the Computers Are Social Actors (CASA) framework indicate that disclosure-based reciprocity can arise even when people interact with artificial agents~\citep{collins1994self, reeves1996media}.
Experimental work also shows reciprocal responding toward computer-based agents, specifically increased user self-disclosure following an agent’s disclosure~\citep{moon2000intimate}.
In studies on conversational AI systems, agent self-disclosure increases users' self-disclosure~\citep{lee2020hear} and improves satisfaction (and reuse intentions) relative to non-disclosing baselines~\citep{park2023effect}.
However, reported benefits are conditional on the contextual relevance and perceived appropriateness of the disclosure~\citep{lee2020hear, park2023effect, tsumura2023influence}.
Notably, prior manipulations of disclosure are typically generic (e.g., emotional vs.\ neutral/factual statements) rather than tailored to the user’s present concern~\citep{lee2017enhancing, ho2018psychological, park2023effect}.

We address this conditionality by operationalizing relevance via concern-aligned persona cues and by comparing no-persona and generic self-disclosure baselines to isolate the added value of contextual alignment.

\subsection{Similarity \& In-Group Cues}
Similarity manipulations affect user judgments and intentions and vary across application domains.
For example, in voice interfaces, matching voice personality increases perceived social presence~\citep{lee2003designing}.
In recommender chatbots, personality alignment improves evaluation and intention measures~\citep{jin2023birds}.
In social robotics, in-group framing yields more favorable judgments (e.g., liking, anthropomorphism) and approach intentions~\citep{nass1997machines, eyssel2012social}.
These operationalizations predominantly rely on static trait/style matching or categorical group labels rather than situational alignment to a user’s ongoing concern.

Taken together, these observations motivate a focus on \emph{concern-aligned} relational cues rather than reliance on generic disclosure or static trait/style matching.

\section{Method}

\label{sec:method}

\subsection{Overview}

The \textbf{Persona Agent} analyzes \textit{dialogue preceding the main conversation} gathered to infer the user’s concerns and related information. Based on the inferred information, it generates a persona that belongs to the user’s in-group and shares the same concerns as the user, embedding this fictional persona--comprising a character profile and curated past experiences--before the dialogue begins. This embedded persona enables the agent to deliver empathetic utterances and self-disclosing narratives, thereby achieving a higher level of rapport compared to chatbots without an explicit, tailored persona. For instance, when responding to a user concerned about work-related stress, a generic chatbot might respond, “I’m sorry you’re feeling stressed; consider taking breaks during work.” In contrast, a persona-embedded agent might share, “I remember feeling overwhelmed at my own job a few years ago. I found that taking short outdoor walks during my breaks helped me manage my stress better.” The latter response leverages persona-driven self-disclosure, which fosters stronger rapport. Such an approach is particularly valuable in domains where user experience is paramount, including psychological support, education, and personalized services. 

\subsection{Persona Definition}

A \textit{persona} in this study consists of a synthetic \textbf{agent background} and \textbf{past narratives}. By injecting this information into the prompt, we create a \textit{persona agent}, a virtual persona capable of interacting with the user. 

\paragraph{Background.}
Background provides a concise overview of who the persona is, including aspects such as personal history, education, or interests. While it does not directly resolve the user’s concern, it frames the agent as a coherent social actor with recognizable traits and contextual grounding. 

\paragraph{Narrative.}
The narrative recounts how the synthetic persona previously faced a challenge similar to the user’s concern and describes the steps they took to overcome it. Based on this synthetic narrative, models engage in strategic self-disclosure that conveys empathy and shared experience.

\subsection{Pre-chat Dialogue Elicitation}

During pre-chat dialogue, a concern-elicitation agent (\textit{collector}) engages the user to elicit the primary concern and brief context. After every two turns, the system checks whether the information is sufficient to proceed; if not, the agent continues with concise clarification questions. Once sufficiency is met, the pipeline advances to persona generation. The fixed sufficiency rule and prompt templates are provided in \Cref{sec:appendix_a}.

\subsection{Persona Generation}

Based on information collected during the pre-chat dialogue phase, 
the system generates \textbf{five in-group personas} that reflect the user’s context and characteristics, 
so that all generated personas would feel relevant and relatable to the user. 

To achieve this, the system categorizes user traits into collected information (e.g., concerns, explicitly stated preferences, or demographic indicators) 
and information unavailable from the pre-chat dialogue (i.e., traits that remained unspecified). 
Based on this categorization, our workflow combines available information traits in different ways to generate five distinct personas. 
Each synthetic persona shares the same primary concern as the user but differs in details of the assigned background and narrative, such as age or profession, to provide diverse perspectives on the shared concern. 
Traits classified as not collected information are left unspecified, so that the resulting profiles only reflect information the user had actually disclosed.

\begin{tcolorbox}[colback=white, colframe=black, arc=3pt, boxrule=0.5pt, width=\linewidth, left=5pt, right=5pt, top=4pt, bottom=4pt]

\textbf{Example of Generated Persona (Abridged)} 

\vspace{4pt} 
\small{
\textbf{Background:} I’m a 26-year-old computer science graduate who developed 
a fascination with AI research during my undergraduate studies, particularly after 
taking courses in machine learning and neural networks\ldots

\vspace{3pt} 

\textbf{Narrative:} After graduating, I was eager to dive into AI research but quickly 
discovered that breaking into the field was more challenging than I anticipated. 
Most research positions required either graduate degrees or significant practical 
experience that I lacked\ldots
}
\end{tcolorbox}
Details of this example can be found in \Cref{sec:appendix_a}(\Cref{lst:ex_prechat,lst:ex_persona}).

\subsection{Persona Quality Validation}
\label{sec:validation}
We evaluate the validity of the \textit{dynamic persona generation} framework through two complementary methods. 

\subsubsection{Persona Evaluation Rubric}
Persona quality is assessed using LLM-based evaluation on two dimensions: \textit{In-group Fitness} and \textit{Concern Resolution Quality}. 
\textit{In-group Fitness} comprises two sub-dimensions, \textit{Shared Background / Identity} \textbf{(IF1)} and \textit{Shared Skills / Interests} \textbf{(IF2)}. 
\textit{Concern Resolution Quality} comprises \textit{Concern Match} \textbf{(CR1)} and \textit{Narrative Authenticity} \textbf{(CR2)}. 
Each sub-dimension is scored on a 0--4 scale, yielding up to 8 points per dimension and a total of 16 points. 

\subsubsection{Pre-chat Dialogue Information Sufficiency Check}
To examine whether the pre-chat dialogue provides adequate information for persona generation, we perform a sufficiency check. 
For a dialogue consisting of $n$ turns, we segment into cumulative intervals of $2, 4, \ldots, n$ turns. 
Personas are generated from each interval, and their quality scores are compared. 
For example, if $n=6$, personas are generated from turns (1--2), (1--4), and (1--6). 

\section{Experiments}
\label{sec:exp}

\subsection{Topic of Conversation}
We set the topic of conversation between the user and the agent to focus on \textbf{\textit{career and employment}}. This domain was chosen because it naturally contains a balanced mix of concerns that require emotional support (e.g., coping with workplace stress) and concerns that require problem-solving or informational guidance (e.g., job search strategies, career planning). This balance makes it a suitable testbed for examining how users interact with a persona-based conversational agent across both affective and cognitive dimensions.

\subsection{Procedure}
Each session consisted of three stages conducted sequentially.(cf. overall architecture in \Cref{fig:main_arch}) All processes, including inference, persona generation, and conversation with the user, were implemented using \textbf{GPT-4o}~\cite{hurst2024gpt}.
\begin{enumerate}
\item \textbf{Pre-chat.} A concern-elicitation agent (\textit{collector}) first engaged the user in a brief conversation to gather the user’s primary concern or topic of discussion. After every two turns of conversation, the system evaluated whether the user’s concern and related contextual information were sufficiently clear. If the information was insufficient, the agent continued to ask concise follow-up questions until enough details were obtained. Once the concern and its related contextual information were sufficiently collected, the session proceeded to the \textit{Profile Selection} stage.
\item \textbf{Profile Selection.} The system generated five persona profiles aligned with the user’s stated concern. Users selected one profile after reviewing textual background information. 
(skipped if \textit{No Persona})
\item \textbf{Post-chat.} The user then engaged in an open-ended chat with the chatbot endowed with the selected persona. After the dialogue, the user completed an online questionnaire.
\end{enumerate}

Throughout the interaction, system and user messages were logged for analysis. The conversation proceeded in a round-robin, turn-based manner, where each turn consisted of a single utterance from either the user or the agent.

\begin{figure*}[!t]
    \centering
    \includegraphics[width=\linewidth]{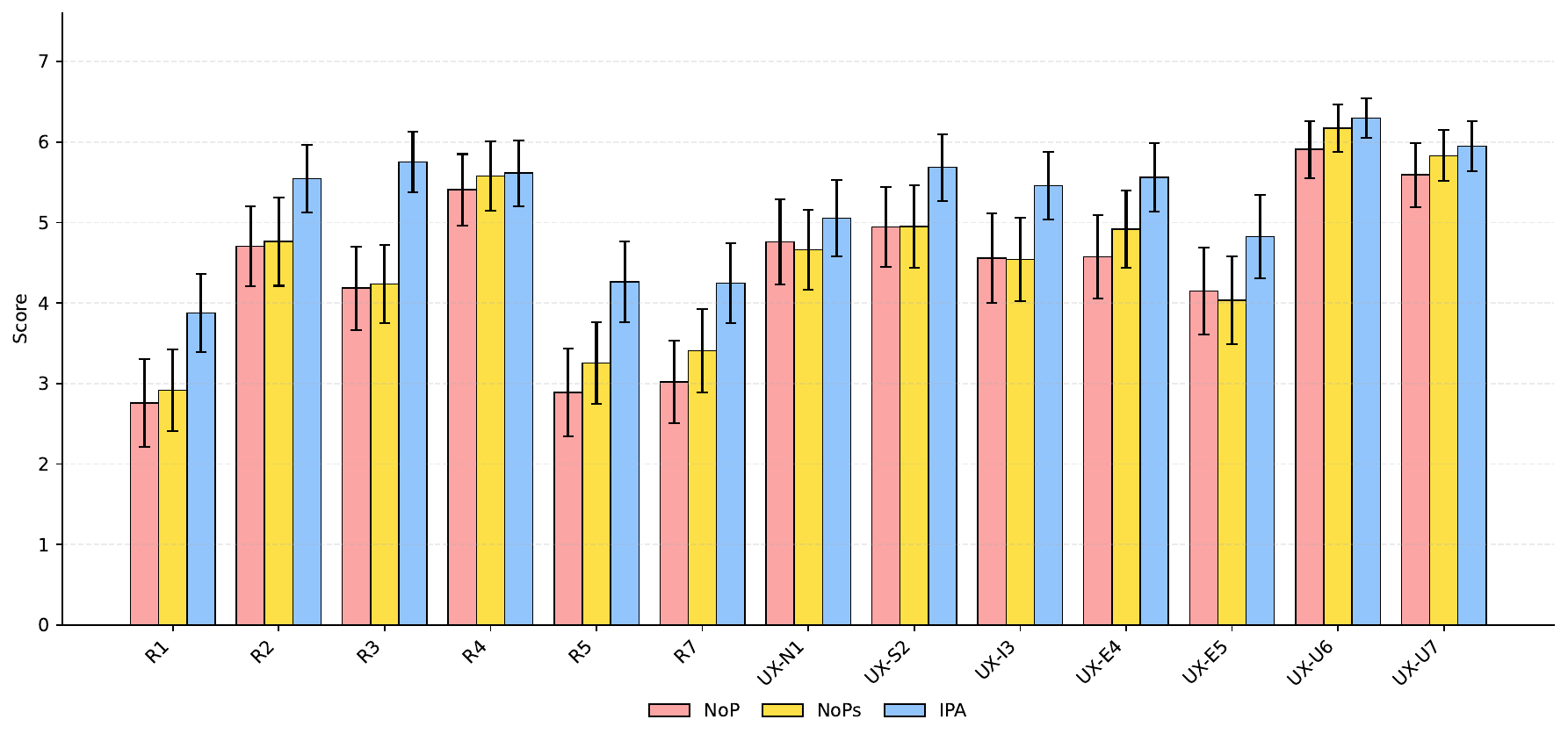}
    \setlength{\abovecaptionskip}{0pt}
    \vspace{-7pt}
    \caption{\textbf{Group comparison of Rapport  and UX items}. Bars display mean $\pm$ 95\% CI for visualization, although all statistical tests were performed on medians due to non-normal distributions (Shapiro–Wilk; see Appendix \ref{sec:detail-stat-analysis}). Rank-based one-sided Mann–Whitney contrasts tested the pair-wise ordered hypothesis \textcolor{Salmon}{NoP} < \textcolor{Dandelion}{NoPs} < \textcolor{Cerulean}{IPA}. Bar colors denote the three conditions: \textcolor{Salmon}{NoP}, \textcolor{Dandelion}{NoPs}, and \textcolor{Cerulean}{IPA}.}
    \label{fig:mw_comparison}
\end{figure*}

\subsection{Participants and Conditions}
We perform a randomized control trial (RCT) to evaluate interactions with assigned persona-conditioned LLM agents, having granted \textit{Institutional Review Board}(IRB) approval. 
All participants provided informed consent prior to participation, and their privacy and anonymity were ensured throughout the study. We recruited 210 participants via CloudResearch’s Connect platform~\cite{hartman2023introducing} and randomly assigned each to one of three experimental conditions ($n=70$ per condition): 

\begin{description}
  \item[\textbf{\emph{Ours:} In-group Persona Agent (IPA)}] \leavevmode\\ 
    Participants conversed with an agent equipped with a user-tailored persona. The system first generated five candidate personas from pre-chat information, and participants selected one before the main dialogue. 
  \item[\textbf{Baseline: No Persona Agent (NoP)}] \leavevmode\\ 
    Participants interacted with an agent that retained dialogue history but had no defined persona. This served as a baseline condition without persona-based framing or self-disclosure.
  \item[\textbf{Baseline: NoP with Self-Disclosure (NoPs)}] \leavevmode\\ 
    Participants conversed with an agent that lacked a persona but could use brief, generic self-disclosing statements (e.g., preferences or feelings) not tailored to the participant’s situation.
\end{description}

Immediately following the interaction, participants completed a post-task questionnaire. One attention-check item instructed participants to choose a specific response. Of the 210 recruited participants, we excluded those who failed the attention check and sessions whose dialogues were off-topic relative to the elicited concern. The final analysis sample comprised 170 participants (\textbf{NoP}: $n{=}54$, \textbf{NoPs}: $n{=}59$, \textbf{IPA}: $n{=}57$).

\subsection{Measures}
Immediately following the interaction, participants completed a post-task questionnaire assessing their perceptions of the agent and the overall dialogue experience. We adopted items from ~\cite{baihaqi2024rapport}, combining measures of \textit{rapport} (six items, R1–R7 except R6\footnote{R6 (``I really care about this virtual agent'') was excluded from the rapport composite due to lexical ambiguity and lack of explicit interaction-referential wording, which reduces sensitivity to single-session effects.}) and \textit{user experience} (seven items). All items were rated on a 7-point Likert scale, with higher scores indicating greater rapport or more positive user experience. The items captured key aspects such as perceived emotional connection, enjoyment, comfort, personal relevance, mutual interest (e.g., ``I feel a bond between this virtual agent and myself'', R5), conversation naturalness and flow, satisfaction, interest, engagement, comprehensibility, and willingness to use the agent again. Details on the questionnaires for rapport and UX are provided in \Cref{sec:appendix_b}.



\subsection{Statistical Analysis}

We analyzed outcomes using rank-based nonparametric methods, given the ordinal response scales and distributional diagnostics. 
Our primary test evaluated the pre-registered ordered alternative \textbf{NoP} $\le$ \textbf{NoPs} $\le$ \textbf{IPA} via a one-sided Jonckheere--Terpstra trend test~\citep{terpstra1952asymptotic,jonckheere1954distribution}. 
To characterize condition differences within each construct family (Rapport or UX), we conducted planned one-sided Mann--Whitney $U$ contrasts for all condition pairs in the hypothesized direction~\citep{mann1947test}, controlling family-wise error within each family using Holm's procedure~\citep{holm1979simple}.
We report effect sizes and descriptive differences alongside adjusted $p$-values; analyses used the preprocessed sample (\textbf{NoP}, $n{=}54$; \textbf{NoPs}, $n{=}59$; \textbf{IPA}, $n{=}57$). 
Further implementation details, assumption checks, and full statistical outputs are provided in \Cref{sec:detail-stat-analysis}.

\section{Results}
\label{sec:result}

\subsection{Main Result}

\paragraph{Overview.}
Group-level comparisons are visualized in \Cref{fig:mw_comparison} and summarized in \Cref{tab:metric_diffs_three_groups}. 
For visualization, data are shown as mean $\pm$ 95\% CI to convey central tendency and variability, but all inferential tests were performed on median ranks using nonparametric procedures due to significant deviations from normality (see \Cref{sec:detail-stat-analysis}).

\paragraph{Rapport.} 
Across all rapport items (R1-R5, R7), scores increased monotonically from \textbf{NoP} to \textbf{NoPs} to \textbf{IPA}, with the largest and most consistent gains for \textbf{IPA} over \textbf{NoP}. 
These improvements were most pronounced on items indexing perceived relatedness and mutual engagement (especially \textbf{R3} ``relevant to me,'' \textbf{R5} ``bond with the agent,'' and \textbf{R7} ``personal interest,'' as well as \textbf{R1} ``think about relationship''). 
In contrast, \textbf{NoPs} provided only modest benefits over \textbf{NoP}, suggesting that generic self-disclosure alone yields limited relational improvement. 
Finally, \textbf{IPA} also exceeded \textbf{NoPs} across all rapport items, indicating that persona alignment adds relational depth beyond disclosure alone.

\paragraph{User Experience.}
Across user-experience items, \textbf{IPA} received the highest scores overall, whereas \textbf{NoP} and \textbf{NoPs} were generally close and did not follow a consistent ordering. 
Relative to \textbf{NoP}, the clearest improvements for \textbf{IPA} centered on \textbf{E4} (engagement) and \textbf{I3} (perceived relevance), alongside more favorable satisfaction and willingness to continue. 
The \textbf{NoPs--NoP} comparison was mixed, with small gains on some items but slight declines on others, indicating that generic self-disclosure did not reliably improve overall experience. 
Finally, \textbf{IPA} exceeded \textbf{NoPs} as well, with the most salient gains again tied to relevance, engagement, and continued-use intent, confirming additional benefits from persona alignment beyond disclosure alone.

\paragraph{Summary.}
Taken together, the analyses reveal a graded pattern of improvement:
(1) \textbf{NoP}~$\rightarrow$~\textbf{NoPs}: generic self-disclosure yields limited and mixed benefits; 
(2) \textbf{NoPs}~$\rightarrow$~\textbf{IPA}: persona alignment adds clear gains, especially on rapport dimensions tied to perceived relevance and personal connection; and 
(3) \textbf{NoP}~$\rightarrow$~\textbf{IPA}: the cumulative difference is most pronounced on key rapport and relevance-related outcomes.
These results show that incorporating concern-aligned in-group persona cues consistently strengthens rapport relative to the baselines, with more modest user-experience gains that are most evident in engagement.

\subsection{Validation Results}
Following the validation procedures described in \Cref{sec:validation}, we conducted two sets of experiments: (1) whether the proposed rubric effectively captures in-group alignment and concern reflection in generated personas; and (2) whether the amount of information collected in the pre-chat dialogue is sufficient for high-quality persona generation.  

\subsubsection{Validation of Persona Quality Rubric}

To examine whether the rubric for persona quality evaluation appropriately captures that a persona generated from the pre-chat dialogue belongs to the user’s in-group and reflects similar concerns, we compared the scores between the group where the persona was evaluated with the pre-chat dia- logue used for its generation (Matched) and the group where the persona was evaluated with a different pre-chat dialogue (Not Matched). \Cref{fig:rubric_comparison}(\Cref{app:pq}) shows that Matched consistently received higher scores across all rubric items, while Not Matched received lower scores. This indicates that evaluation with the rubric can effectively verify whether a persona is appropriate for the user.

\subsubsection{Validation of Pre-chat Dialogue Sufficiency Check}
In the pre-chat dialogue stage, we verify whether the information collected every two turns is sufficient for persona generation. To examine this, a subset of 70 pre-chat dialogues from the full set collected in the experiment was segmented into cumulative intervals of two turns, and a persona was generated at each interval. In total, 480 personas were generated and subsequently evaluated for quality.
Overall, the results demonstrate that the rubric-based evaluation is sensitive to information sufficiency, assigning lower scores to personas derived from incomplete contexts and higher scores to those generated from sufficient dialogue history. Detailed results are provided in \Cref{sec:appendix_suff_table}.

\section{Analysis}
\label{sec:analysis}

\subsection{Correlation Between Persona Quality and Rapport Score}
Following the validation of the persona quality rubric in the previous analysis, we conducted correlation analyses to examine whether rubric scores were associated with self-reported outcomes (rapport and UX; see \Cref{tab:rubric_score_distribution,tab:rubric_raport_pearson} for detailed results). The selected personas generally demonstrated high rubric scores, with Narrative Authenticity showing zero variance and thus being excluded from correlation interpretation.

Pearson correlation analysis revealed that overall persona quality (Total) exhibited significant positive correlations with R5 (sense of connection, $r=0.36, p<.05$) among rapport items and notably with E5 (intention to continue conversation in the future, $r=0.37, p<.05$) among UX items. At the dimensional level, Shared Skills showed the strongest correlation with E5 ($r=0.48, p<.05$), linking shared interests to continued use. In contrast, no consistent significant correlations emerged across other rapport items, indicating that persona quality is selectively associated with specific outcomes such as ``sense of connection'' and ``continuation intention'' rather than comprehensively explaining rapport as a whole.

\subsection{Turn-level Behavioral Complement.} While self-reported rapport captures the overall experience, it fails to reveal \emph{specific} behavioral nuances across conditions. To address this, we conducted a turn-level analysis focusing on \emph{immediate partner responses} within the dialogue logs. Specifically, we scored each utterance using an LLM rubric (0--3) for self-disclosure ($sd_t$) and empathy ($emp_t$), and subsequently classified them as ``high'' or ``low'' based on fixed thresholds (details in \Cref{sec:app_llm_judging,sec:app_metrics}). We then calculated the conditional probability of a response in speaker-change adjacent pairs, conditioned on the partner's preceding behavior. We report these probabilities along with their difference, defined as the reciprocity index $R$ (\Cref{tab:app_reciprocity_post}).

\begin{table}[t]
\centering
\scriptsize
\setlength{\tabcolsep}{1pt}
\begin{tabular}{lccccccc}
\toprule
Condition & {$p^{A\to U}_{sd|high}$} & {$p^{A\to U}_{sd|low}$} & \textbf{{$R^{A\to U}_{SD}$}} & {$p^{U\to A}_{emp|high}$} & {$p^{U\to A}_{emp|low}$} & \textbf{{$R^{U\to A}_{EMP}$}} \\
\midrule
NoP  & \texttt{NaN} & 0.377 & \textbf{\texttt{NaN}} & 0.908 & 0.631 & \textbf{+0.277} \\
NoPs  & 0.425 & 0.454 & \textbf{-0.029} & 0.993 & 0.936 & \textbf{+0.057} \\
IPA  & 0.620 & 0.249 & \textbf{+0.371} & 0.903 & 0.539 & \textbf{+0.364} \\
\bottomrule
\end{tabular}
\caption{\textbf{Directional reciprocity in the \texttt{post} segment.}
Each $p$ denotes a conditional probability of a next-turn response event (User deep self-disclosure or Agent empathy) given whether the immediately preceding turn contains deep self-disclosure (event present vs.\ absent; ``high/low''). $R$ is the difference of the two conditional probabilities (e.g., $R^{A\to U}_{SD}=p^{A\to U}_{sd\mid high}-p^{A\to U}_{sd\mid low}$). \texttt{NaN} indicates non-estimability due to event sparsity.}
\label{tab:app_reciprocity_post}
\end{table}


We analyze the \texttt{post} segment to capture interactions after persona manipulation was active (see \Cref{sec:app_reciprocity}). IPA shows the clearest User self-disclosure reciprocity in response to the agent ($A \to U$): deep User self-disclosure was more likely after deep agent self-disclosure than after non-deep agent turns ($0.620$ vs. $0.249$; $R^{A\to U}_{SD}=+0.371$). Agent empathy following User self-disclosure ($U \to A$) was positive across conditions, with the largest increase in IPA ($R^{U\to A}_{EMP}=+0.364$), while NoPs showed a smaller increase partly due to its near-ceiling baseline empathy. 

Overall, these results suggest that IPA uniquely strengthens self-disclosure reciprocity, while the agent's empathic responsiveness to User self-disclosure is broadly present across conditions.

\section{Conclusion}
This work demonstrates that embedding in-group personas into conversational AI can meaningfully enhance rapport and user experience. By combining pre-chat dialogue elicitation with structured persona framing and self-disclosure, our approach enables agents to appear more relatable, empathetic, and engaging. 
After pre-processing, we analyzed data from 170 participants and found that persona-embedded agents consistently outperformed both non-persona and self-disclosing baselines on rapport measures, including perceived personal relevance.
Overall, lightweight, prompt-based persona design offers a practical means to strengthen rapport and engagement in supportive interpersonal settings--supported by analyses linking persona quality to users’ sense of connection and continuation intention, and turn-level evidence of increased self-disclosure reciprocity--and extending this framework to more sensitive domains remains an important direction for future work.

\section{Limitations}
\label{sec:limit}
Although our study offers insights into the role of in-group personas in conversational AI, it also carries several limitations that should be acknowledged. 

\begin{description}
  \item[Anthropomorphism:]While persona-driven design can enhance users' sense of connection, excessive anthropomorphic cues may also introduce unintended side effects. Overly human-like behaviors, such as emotional overexpression or self-referential remarks, can create unrealistic expectations about the agent's capabilities or authenticity. This mismatch between user perception and system intent may lead to confusion or reduced trust, suggesting that careful calibration of anthropomorphic elements is essential for maintaining credible and comfortable human--AI interaction.

  \item[Pre-chat Sufficiency:] Our analyses focus on sessions that met a minimum level of pre-chat sufficiency, enabling personas to be grounded in user-provided cues. Because we did not systematically collect a comparable set of ``insufficient'' pre-chat cases, we cannot conduct a reliable ablation on how pre-chat quantity affects persona quality and downstream outcomes. This limits our understanding of robustness under sparse inputs, suggesting that future work should evaluate broader ranges of pre-chat completeness.

  \item[Disclosure-Averse Users:] Our framework assumes that users are comfortable articulating their primary concern during the pre-chat stage, and our experiments were designed around participants for whom such disclosure is a relatively low burden. In practice, some users may be hesitant to share personal details upfront, potentially reducing usability and limiting persona quality. While the pre-chat agent is intended to ease disclosure through guided prompting, disclosure-averse populations remain out of scope for the current study, motivating future work in this direction.

  \item[Persona Selection Effect:]In the IPA condition, participants selected one persona from five generated candidates. We adopted this design because the brief pre-chat phase may not provide enough information to reliably identify a single best-matching persona automatically, and because allowing users to review and choose a disclosed persona can help them select the perspective they personally need in a peer-support-like setting. However, this choice may also have increased perceived relevance, rapport, or satisfaction, potentially overstating the effect of the in-group persona itself.

\end{description}

Future work should address these limitations by expanding the scope of participants and analyses in order to provide more generalizable insights.




\section{Ethical Considerations}
\label{sec:ethics}

\paragraph{Potential Harms of Synthetic In-group Personas.}
We acknowledge that in sensitive domains such as mental health, trauma recovery, or medical advice, a synthetic persona that feigns shared human experience can lead to severe consequences. In such contexts, ``synthetic empathy'' may induce undue trust, discourage seeking professional help, or otherwise exacerbate vulnerability. Our study, however, was conducted in the comparatively lower-risk context of career and employment counseling, rather than in high-stakes settings such as suicide prevention or grief counseling. The persona's ``experience'' was limited to professional challenges and workload management, and the interactions were framed as instrumental support rather than therapeutic intervention. Accordingly, we expect the immediate risk of harm to be substantially lower than in clinical settings, while noting that similar risks could arise if such personas were deployed in more sensitive contexts.

\paragraph{Participant Disclosure and Right to Withdraw.}
Prior to participation, we informed participants in the study introduction about potential risks and disadvantages, including the possibility of discomfort during experimental procedures. Participants were explicitly informed that they could withdraw at any time without penalty. We also indicated that appropriate support would be available if any discomfort persisted. We provide the disclosure text in the appendix, with selected portions redacted to preserve the authors' anonymity.

\paragraph{Scope and Limitations.}
These ethical considerations apply to the specific experimental scope described above. We caution against generalizing the safety of synthetic in-group personas to high-stakes or clinical domains without domain-specific safeguards, oversight, and evaluation.


\bibliography{custom, anthology}

\appendix
\section*{Appendix}
\label{sec:appendix}

\section{Experiment Detail}
\label{sec:ext-detail}

This section includes details of our experiment. \Cref{tab:metric_diffs_three_groups} shows overall result of human evaluation.

\subsection{Demographic Distributions of Participants}
\begin{table}[ht]
\centering
\small
\setlength{\tabcolsep}{6pt}
\begin{tabular}{lccc}
\toprule
 & \textbf{Category} & \textbf{n (\%)} &  \\
\midrule
\multicolumn{4}{l}{\textbf{Gender}} \\
 & Man & 86 (50.6\%) & \\
 & Woman & 78 (45.9\%) & \\
 & Missing & 3 (1.8\%) & \\
 & Prefer not to say & 2 (1.2\%) & \\
 & Agender & 1 (0.6\%) & \\
\midrule
\multicolumn{4}{l}{\textbf{Race}} \\
 & White & 127 (74.7\%) & \\
 & Black or African American & 14 (8.2\%) & \\
 & An ethnicity not listed here & 7 (4.1\%) & \\
 & Chinese & 6 (3.5\%) & \\
 & Vietnamese & 4 (2.4\%) & \\
 & Missing & 3 (1.8\%) & \\
 & Filipino & 3 (1.8\%) & \\
 & Asian Indian & 2 (1.2\%) & \\
 & Japanese & 1 (0.6\%) & \\
 & Other & 1 (0.6\%) & \\
 & Korean & 1 (0.6\%) & \\
 & American Indian or Alaska Native & 1 (0.6\%) & \\
\midrule
\multicolumn{4}{l}{\textbf{Employment Status}} \\
 & Full-time & 80 (47.1\%) & \\
 & Part-time & 27 (15.9\%) & \\
 & Unemployed & 18 (10.6\%) & \\
 & Not in paid work & 14 (8.2\%) & \\
 & Business Owner & 10 (5.9\%) & \\
 & Student & 9 (5.3\%) & \\
 & Retired & 6 (3.5\%) & \\
 & Prefer not to say & 3 (1.8\%) & \\
 & Missing & 3 (1.8\%) & \\
\bottomrule
\end{tabular}
\caption{\textbf{Demographic distributions (n, \%).} Total 170 participants, missing 3 informations.}
\label{tab:demographics_big}
\end{table}

\Cref{tab:demographics_big} shows demographic distributions of the participants. Overall, the sample exhibits reasonable variation in background (e.g., gender and employment situations), and participants engaged in conversations about their own real career and employment concerns rather than role-played or hypothetical scenarios. At the same time, the pool is geographically and linguistically constrained (U.S.-based, English-speaking) and was recruited from a single crowdsourcing platform, which may limit broader generalizability.

\subsection{Participant Disclosure: Risks, Withdrawal, and Support}
\label{app:disclosure}
Participants were informed of this procedure in advance as part of the study instructions and consent process. They were also informed of potential risks or discomforts, data handling and privacy protections, and that participation was voluntary and could be withdrawn at any time without penalty.
\begin{tcolorbox}[colback=white, colframe=black, arc=3pt, boxrule=0.5pt, width=\linewidth, left=5pt, right=5pt, top=4pt, bottom=4pt]

\textbf{Disadvantages, Risks of Participation, and Compensation for Harm}

\vspace{4pt}
\small{
Since the study involves discussing personal concerns with a chatbot, psychological discomfort or risks may arise. You are free to stop the experiment at any time without notifying the researcher. If you experience psychological discomfort after completion or withdrawal, you may contact the principal investigator listed below, who will consult with the university(Korea University) to arrange appropriate support.
}
\end{tcolorbox}

\subsection{Differences Between IPA and Baselines}
\label{app:ipa_vs_baselines}

All three conditions used the same user messages as conversational context, ensuring that the baseline personalization naturally provided by LLMs was held constant across conditions. The only difference was the type of additional personalization provided by the agent: \textbf{NoP} included no added personalization cues, \textbf{NoPs} included a brief generic self-disclosure, and \textbf{IPA} incorporated a concern-aligned in-group persona derived from pre-chat information. Therefore, the observed advantages of \textbf{IPA} are attributable to the added in-group persona cues rather than generic personalization. Moreover, \textbf{IPA} produced the strongest gains specifically in relevance and perceived personal connection, which are theoretically linked to perceived in-group similarity. This design thus isolates the contribution of in-group persona alignment.

\paragraph{Example of Actual Agent Responses.}
\begin{description}
  \item[\textbf{\emph{Ours:} In-group Persona Agent (IPA)}] \leavevmode\\
    \textit{(generic personalization + self-disclosure with persona information)}\\
    Agent: Yes, after realizing my efforts weren't fully visible, I increased my visibility with upper management and successfully secured a promotion along with a revised compensation package.

  \item[\textbf{Baseline: No Persona Agent (NoPs)}] \leavevmode\\
    \textit{(generic personalization + self-disclosure without persona information)}\\
    Agent: That's perfectly fine! Medical school will expose you to a variety of specialties, which might help clarify your interests. I admire cardiology because of the heart's intricate systems, but every specialty has its unique appeal. What aspects of medicine spark your curiosity?

  \item[\textbf{Baseline: NoP (NoP)}] \leavevmode\\
    \textit{(generic personalization only)}\\
    Agent: That's a great goal! To transition, build your technical skills, particularly in areas like network architecture, encryption, and coding. Seek out projects or training that give you engineering experience, and consider networking with professionals in the field for mentorship and guidance.
\end{description}

\subsection{Participant Openness and Ecological Validity of a Paid Sample}
Our framework targets users who intentionally seek concern-oriented, personal conversations; thus, even though participants were compensated, they voluntarily engaged with a framework that clearly required sharing personal context, making the sample reasonably comparable to real-world users who would opt in. To reduce the burden of self-disclosure, we use a lightweight pre-chat agent that starts from the user’s stated concern and asks brief, context-relevant follow-up questions, framing onboarding as collaborative problem identification rather than form filling. We also acknowledge that some users are highly disclosure-averse; future work will explore methods to assess disclosure comfort early and offer alternative onboarding strategies tailored to different willingness levels.

\begin{table*}[ht]
  \centering
  \renewcommand{\arraystretch}{0.9}
  \begin{tabular}{l|ccc|ccc|ccc}
    \toprule
    \textbf{Metric} &
    \textbf{IPA} & \textbf{NoP} & \textbf{NoPs} &
    $\Delta_{\text{IPA,NoP}}$ & $\Delta_{\text{IPA,NoPs}}$ & $\Delta_{\text{NoPs,NoP}}$ &
    $r_{\text{IPA,NoP}}$ & $r_{\text{IPA,NoPs}}$ & $r_{\text{NoPs,NoP}}$ \\
  \midrule
    R1 & \textbf{3.88} & 2.76 & 2.92 & \cellcolor{gray!20}\textbf{1.12} & \cellcolor{gray!20}\textbf{0.96} & 0.16 & 0.29 & 0.26 & 0.05 \\
    R2 & \textbf{5.54} & 4.70 & 4.76 & \textbf{0.84} & 0.78 & 0.06 & 0.23 & 0.17 & 0.04 \\
    R3 & \textbf{5.75} & 4.19 & 4.24 & \cellcolor{gray!20}\textbf{1.57} & \cellcolor{gray!20}\textbf{1.52} & 0.05 & \textbf{0.42} & \textbf{0.42} & 0.01 \\
    R4 & \textbf{5.61} & 5.41 & 5.58 & 0.21 & 0.04 & 0.17 & 0.06 & -0.01 & 0.06 \\
    R5 & \textbf{4.26} & 2.89 & 3.25 & \cellcolor{gray!20}\textbf{1.37} & \cellcolor{gray!20}\textbf{1.01} & 0.37 & \textbf{0.34} & 0.25 & 0.10 \\
    R7 & \textbf{4.25} & 3.02 & 3.41 & \cellcolor{gray!20}\textbf{1.23} & \textbf{0.84} & 0.39 & \textbf{0.30} & 0.21 & 0.10 \\
    UX-N1 & \textbf{5.05} & 4.76 & 4.66 & 0.29 & 0.39 & -0.10 & 0.08 & 0.11 & -0.03 \\
    UX-S2 & \textbf{5.68} & 4.94 & 4.95 & \textbf{0.74} & \textbf{0.74} & 0.00 & 0.20 & 0.19 & 0.02 \\
    UX-I3 & \textbf{5.46} & 4.56 & 4.54 & \textbf{0.90} & \textbf{0.91} & -0.01 & 0.21 & 0.23 & -0.01 \\
    UX-E4 & \textbf{5.56} & 4.57 & 4.92 & \cellcolor{gray!20}\textbf{0.99} & \textbf{0.65} & 0.34 & 0.27 & 0.18 & 0.09 \\
    UX-E5 & \textbf{4.82} & 4.15 & 4.03 & 0.68 & 0.79 & -0.11 & 0.17 & 0.19 & -0.03 \\
    UX-U6 & \textbf{6.30} & 5.91 & 6.17 & 0.39 & 0.13 & 0.26 & 0.14 & 0.04 & 0.10 \\
    UX-U7 & \textbf{5.95} & 5.59 & 5.83 & 0.35 & 0.12 & 0.24 & 0.10 & 0.05 & 0.06 \\
  \bottomrule
  \end{tabular}
       \caption{
    \textbf{Human Evaluation Results.}
    Group means, pairwise mean differences ($\Delta$), and effect sizes ($r=Z/\sqrt{N}$). 
    \textbf{Bold} indicates statistically significant differences ($p<.05$) or medium effects ($r\ge.30$). 
    \cellcolor{gray!20}\textbf{Shaded} indicates strong significance ($p<.01$). 
    Pairwise $p$-values from one-sided Mann--Whitney tests, Holm-adjusted within item. 
    Effect-size benchmarks: small ($.10\!\le\!r\!<\!.30$), medium ($.30\!\le\!r\!<\!.50$).
    }
  \label{tab:metric_diffs_three_groups}
\end{table*}

\section{Detail Statistical Analysis}
\label{sec:detail-stat-analysis}

\paragraph{Implementation Details.}
Analyses were conducted in Python 3.13.7 using NumPy 2.3.3, Pandas 2.3.3, SciPy 1.16.2, Pingouin 0.5.5, and Matplotlib 3.10.6.
The rpy2 3.6.4 interface connected to R 4.3.1 with the clinfun 1.0.16 package for Jonckheere--Terpstra tests.

\paragraph{Assumption Testing.}
Distributional assumptions were formally evaluated for each item to verify the appropriateness of parametric inference. 
Shapiro--Wilk tests~\citep{shapiro1965analysis} revealed significant deviations from normality across most conditions. Levene's tests~\citep{levene1960robust} suggested that variance heterogeneity was not pervasive, although a small number of items showed evidence of heteroscedasticity (minimum $p = .013$). Given the ordinal scales and these diagnostics, we used rank-based nonparametric analyses. 
\Cref{tab:assumption-tests} summarizes these results.

\paragraph{Analytic Framework.}
Given the ordinal response scale and the non-normal distributions observed, rank-based nonparametric analyses were adopted. 
The Jonckheere--Terpstra trend test~\citep{terpstra1952asymptotic,jonckheere1954distribution} was used to evaluate the pre-specified ordered alternative (\textbf{NoP} $\le$ \textbf{NoPs} $\le$ \textbf{IPA}) in a one-sided manner. 
When a significant monotonic trend was detected, planned one-sided pairwise Mann--Whitney $U$ tests~\citep{mann1947test} were performed to decompose the pattern in the hypothesized direction.

Family-wise error was controlled separately within each construct family: Rapport (R1--R5, R7) and UX (N1, S2, I3, E4, E5, U6, U7). 
Within each family, Holm's sequentially rejective method~\citep{holm1979simple} was applied, and adjusted $p$-values ($p_{\text{adj}}$) are reported for each item. 
Statistical significance was defined as $p_{\text{adj}}<.05$ (one-sided). 
All analyses were conducted at $\alpha=0.05$ on the preprocessed sample (\textbf{NoP}, $n{=}54$; \textbf{NoPs}, $n{=}59$; \textbf{IPA}, $n{=}57$).

\paragraph{Reporting Conventions and Effect Sizes.}
For Mann--Whitney pairwise contrasts, effect sizes were computed as $r = Z/\sqrt{N}$ following Rosenthal's $r$-equivalent formulation~\citep{rosenthal1991meta}. 
Magnitudes were interpreted analogously to correlation benchmarks~\citep{cohen1992power}: small ($r\approx .10$) and medium ($r\approx .30$). 
Across both trend-level and pairwise analyses, $p$-values are treated as inferential thresholds, while raw differences ($\Delta$) are reported as descriptive complements; substantive interpretation emphasizes the direction and magnitude of effects.

\begin{table}[t]
\centering

\scriptsize

\begin{subtable}[t]{0.9\columnwidth}
\centering
\subcaption{Shapiro--Wilk tests of normality by condition.}
\begin{tabular}{l l c c c}
\toprule
\textbf{Item} & \textbf{Group} & \textbf{N} & \textbf{W} & \textbf{$p$} \\
\midrule
R1 & NoP & 54 & 0.816 & <.001 \\
R1 & NoPs & 59 & 0.859 & <.001 \\
R1 & Peer & 57 & 0.921 & .0011 \\

R2 & NoP & 54 & 0.906 & <.001 \\
R2 & NoPs & 59 & 0.861 & <.001 \\
R2 & Peer & 57 & 0.802 & <.001 \\

R3 & NoP & 54 & 0.928 & .0030 \\
R3 & NoPs & 59 & 0.928 & .0018 \\
R3 & Peer & 57 & 0.809 & <.001 \\

R4 & NoP & 54 & 0.859 & <.001 \\
R4 & NoPs & 59 & 0.808 & <.001 \\
R4 & Peer & 57 & 0.825 & <.001 \\

R5 & NoP & 54 & 0.843 & <.001 \\
R5 & NoPs & 59 & 0.892 & <.001 \\
R5 & Peer & 57 & 0.929 & .0024 \\

U7 & NoPs & 59 & 0.816 & <.001 \\
U7 & Peer & 57 & 0.821 & <.001 \\
\multicolumn{5}{c}{\textit{(Other items show comparable non-normal patterns.)}} \\
\bottomrule
\end{tabular}
\end{subtable}

\par\medskip

\begin{subtable}[t]{0.9\columnwidth}
\centering
\subcaption{Levene's tests of homogeneity of variance across groups.}
\begin{tabular}{l c c c c}
\toprule
\textbf{Item} & \textbf{$k$} & \textbf{Center} & \textbf{$F$} & \textbf{$p$} \\
\midrule
R1 & 3 & Mean & 1.63 & .200 \\
R2 & 3 & Mean & 3.95 & .021 \\
R3 & 3 & Mean & 4.43 & .013 \\
R4 & 3 & Mean & 0.30 & .738 \\
R5 & 3 & Mean & 0.08 & .920 \\
R7 & 3 & Mean & 0.35 & .708 \\
E4 & 3 & Mean & 2.00 & .139 \\
E5 & 3 & Mean & 0.20 & .820 \\
I3 & 3 & Mean & 3.61 & .029 \\
N1 & 3 & Mean & 0.23 & .797 \\
S2 & 3 & Mean & 2.17 & .117 \\
U6 & 3 & Mean & 1.56 & .213 \\
U7 & 3 & Mean & 3.06 & .050 \\
\bottomrule
\end{tabular}
\end{subtable}

\caption{\textbf{Tests of distributional assumptions for all items.}
Shapiro--Wilk tests assessed normality within each group, and
Levene's tests examined homogeneity of variance across conditions.}
\label{tab:assumption-tests}
\end{table}

\begin{table}[h]
\setlength{\tabcolsep}{3pt}
\centering
\begin{tabular}{lcccc}
\toprule
Item & $J_T$ & $p$ & $p_{\text{adj}}$ & $r$ \\
\midrule
R1 & 5921.5 & .001 & .002 & 0.243 \\
R2 & 5633.5 & .009 & .019 & 0.180 \\
R3 & 6404.5 & <.001 & <.001 & 0.349 \\
R4 & 5017.5 & .280 & .280 & 0.045 \\
R5 & 6095.0 & <.001 & .001 & 0.281 \\
R7 & 5948.0 & .001 & .002 & 0.249 \\
\midrule
N1 & 5101.0 & .205 & .249 & 0.063 \\
S2 & 5571.0 & .015 & .075 & 0.166 \\
I3 & 5621.0 & .010 & .062 & 0.177 \\
E4 & 5809.0 & .002 & .015 & 0.219 \\
E5 & 5442.5 & .036 & .144 & 0.138 \\
U6 & 5316.5 & .075 & .225 & 0.110 \\
U7 & 5216.5 & .124 & .249 & 0.088 \\
\bottomrule
\end{tabular}
\caption{\textbf{Jonckheere--Terpstra trend test results for both Rapport and User Experience constructs.} Reported are test statistics ($J_T$), one-sided $p$-values, Holm-adjusted $p$-values (within construct), and effect sizes ($r=Z/\sqrt{N}$).}
\label{tab:jt_combined}
\end{table}

\paragraph{Trend-level Inference.}
Jonckheere--Terpstra tests examined the pre-registered ordered alternative 
(\textbf{NoP} $\le$ \textbf{NoPs} $\le$ \textbf{IPA}) across all items within each construct. 
As shown in~\Cref{tab:jt_combined}, the monotonic trend was significant for most \textbf{Rapport} items, providing evidence for a stepwise increase in relational judgments across conditions. 
For rapport items (R1--R7 except R6), five of six reached significance after Holm adjustment ($p_{\text{adj}} \le .019$), with small-to-medium effects ($r = .18$--.35). 
Notably, \textbf{R3} (“This virtual agent is very relevant to me”) exhibited the strongest trend ($r = .349$, $p_{\text{adj}} < .001$), followed by \textbf{R5} ($r = .281$, $p_{\text{adj}} = .001$), suggesting that identity-aligned persona framing most strongly shaped perceived personal relevance and connection. 
In contrast, \textbf{R4} showed no reliable monotonic trend ($p_{\text{adj}} = .28$, $r = .045$), indicating an item-specific exception to the ordered pattern.

User-experience items (N1, S2, E4, E5, I3, U6, U7) displayed weaker and less consistent ordered evidence. 
Only \textbf{E4} (“engagement”) remained significant after correction ($p_{\text{adj}} = .015$, $r = .219$), while the remaining UX items did not reach adjusted significance ($p_{\text{adj}} \ge .062$; $r \le .177$). 
Together, these JT results indicate that the pre-registered ordered improvement is robust for \textbf{Rapport}, whereas for \textbf{User Experience} the trend is more modest and primarily driven by engagement.

\paragraph{Pairwise Contrasts.}
Building on the significant ordered trends, planned one-sided Mann--Whitney $U$ tests compared all condition pairs (\textbf{NoP}$\le$\textbf{NoPs}, \textbf{NoPs}$\le$\textbf{IPA}, \textbf{NoP}$\le$\textbf{IPA}). 
Detailed pairwise results, including adjusted $p$-values, mean differences ($\Delta$), and effect sizes ($r$), are summarized in Table~\ref{tab:metric_diffs_three_groups}. 
Interpretation emphasizes the magnitude and direction of these effects, with $p$-values serving only as inferential filters and $\Delta$ as descriptive complements.

\begin{table}[ht]
\centering
\scriptsize
\begin{tabular}{lccc}
\hline
Score Dimension & Mean & Variance & Scale \\
\hline
Total             & 14.43 & 1.09 & 0--16 \\
Shared Background & 3.26  & 0.44 & 0--4  \\
Shared Skills     & 3.21  & 0.60 & 0--4  \\
Concern Match     & 3.85  & 0.57 & 0--4  \\
Narrative Auth    & 4.00  & 0.00 & 0--4  \\
\hline
\end{tabular}
\caption{\textbf{Rubric Score Distribution (Chosen Personas)}}
\label{tab:rubric_score_distribution}
\end{table}

\begin{table}[ht]
\centering
\scriptsize
\setlength{\tabcolsep}{2pt}
\begin{tabular}{lcccc}
\hline
Item 
& Total Score
& Shared Background 
& Shared Skills 
& Concern Match \\
\hline
R1 & 0.211 & 0.091 & 0.212 & 0.108 \\
R2 & 0.209 & 0.229 & 0.270 & -0.067 \\
R3 & 0.202 & 0.111 & 0.278 & 0.004 \\
R4 & 0.321 & 0.274 & 0.298 & 0.082 \\
R5 & \textbf{0.364} & 0.190 & 0.328 & 0.199 \\
R7 & 0.256 & 0.074 & 0.115 & 0.310 \\
\midrule
N1 & \textbf{0.352} & 0.291 & 0.277 & 0.151 \\
S2 & 0.230 & 0.200 & 0.301 & -0.036 \\
I3 & \textbf{0.345} & 0.266 & 0.248 & 0.189 \\
E4 & 0.249 & 0.174 & 0.203 & 0.124 \\
E5 & \textbf{0.371} & 0.304 & \textbf{0.483} & -0.042 \\
U6 & 0.045 & 0.014 & 0.025 & 0.049 \\
U7 & 0.169 & 0.096 & 0.174 & 0.062 \\
\hline
\end{tabular}
\caption{
\textbf{Post-wise Similarity Scores.}
\textbf{Bold} indicates statistically significant Pearson correlations ($p<.05$).
}
\label{tab:rubric_raport_pearson}
\end{table}

\section{Persona Quality: Matched vs. Unmatched}
\label{app:pq}

\begin{figure}[h]
    \centering
    \includegraphics[width=0.9\linewidth]{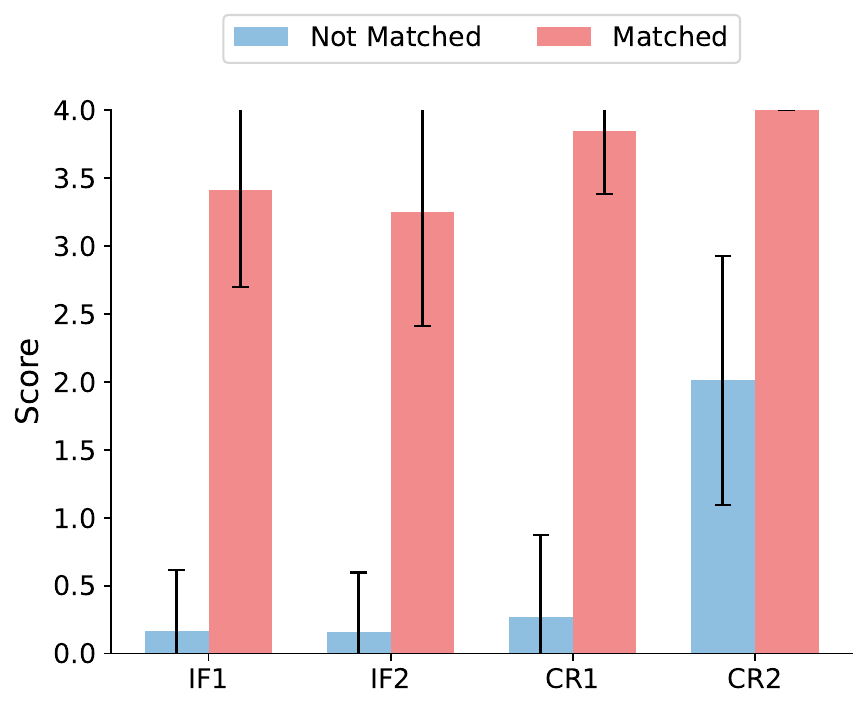}
    \caption{\textbf{Rubric Score comparison} between \textit{Matched} and \textit{Not Matched}. Bar colors denote the condition \textcolor{Cerulean}{Not Matched} and \textcolor{Salmon}{Matched}.}
    \label{fig:rubric_comparison}
    \vspace{-0.5em}
\end{figure}

\Cref{fig:rubric_comparison} reports item-level persona-quality scores for the Matched and Not Matched settings. Across all rubric items, the Matched setting yields higher scores (3.42, 3.25, 3.85, 4.00), whereas the Not Matched setting yields lower scores (0.17, 0.16, 0.27, 2.01). This consistent separation indicates that the rubric can help verify whether a persona is appropriate for the user given the corresponding pre-chat dialogue.

\section{Evaluation Scores by Pre-chat Dialogue Length.}
\label{sec:appendix_suff_table}

Each row indicates the number of turns used for persona generation, denoted as ``Original Turn--Used Turn'' (e.g., 4-2 means a pre-chat dialogue of original length 4, using its first 2 turns). Scores are reported as Mean (Std). \textit{Num} indicates the number of generated personas. Within each Max Turn block, the highest value in each column is shown in bold. All evaluations were conducted with a temperature of 0, while all persona generations were performed with a temperature of 1.

As shown in \Cref{tab:suff_table}, personas generated from dialogues with a longer maximum turn length tend to receive lower scores when evaluated at earlier segments (e.g., 6-2), reflecting the fact that longer pre-chat dialogues arise when the initial exchanges are insufficient. Consequently, personas generated from the first few turns of such dialogues lack sufficient information and are judged to be of lower quality. In contrast, when the full set of turns is used (e.g., 6-6), the scores improve substantially.

Importantly, the differences are particularly pronounced in the Ingroup-fitness dimensions (IF1 and IF2): insufficient dialogue context primarily affects the model’s ability to generate personas that align with the user’s group identity and shared concerns. By contrast, the consistency-related criteria (CR1 and CR2) remain relatively stable across conditions. These results demonstrate that the rubric-based evaluation is sensitive to information sufficiency, assigning lower scores to personas derived from incomplete contexts and higher scores to those generated from sufficient dialogue history.

\begin{table}[t]
\centering
\resizebox{\columnwidth}{!}{
\begin{tabular}{c|c|c|c|c|c|c}
\toprule
Turn & Num & IF1 & IF2 & CR1 & CR2 & Sum \\
\midrule
2-2 & 109 & 3.25 (0.65) & 3.06 (0.87) & 3.99 (0.10) & 4.00 (0.00) & 14.30 (1.38) \\
\midrule
4-2 & 109 & 2.88 (0.97) & 2.61 (0.90) & 3.96 (0.19) & \textbf{4.00 (0.00)} & 13.46 (1.72) \\
4-4 & 187 & \textbf{3.19 (0.64)} & \textbf{3.17 (0.85)} & \textbf{3.98 (0.14)} & \textbf{4.00 (0.00)} & \textbf{14.33 (1.30)} \\
\midrule
6-2 & 9   & 1.89 (1.37) & 1.22 (1.13) & 3.11 (1.29) & 3.89 (0.31) & 10.11 (3.38) \\
6-4 & 31  & 2.52 (0.95) & 2.65 (0.78) & \textbf{3.84 (0.37)} & \textbf{4.00 (0.00)} & 13.00 (1.83) \\
6-6 & 35  & \textbf{2.94 (0.75)} & \textbf{2.86 (0.76)} & 3.80 (0.52) & \textbf{4.00 (0.00)} & \textbf{13.60 (1.74)} \\
\bottomrule
\end{tabular}
}
\caption{\textbf{Evaluation Scores by Pre-chat Dialogue Length.} }
\label{tab:suff_table}
\vspace{-0.5em}
\end{table}

\section{Turn-level Behavioral Annotation and Reciprocity Analysis}
\label{sec:app_turnlevel_overview}

\subsection{LLM Rubric Judging for Self-Disclosure and Empathy}
\label{sec:app_llm_judging}
\paragraph{Judging Unit.}
Each turn is judged \emph{independently} (single-utterance judging), producing two scores:
self-disclosure depth $sd_t\in\{0,1,2,3\}$ and empathy level $emp_t\in\{0,1,2,3\}$.
The LLM is constrained to return a JSON object containing scores and short rationales. Details of this example can be found in \Cref{lst:llm_judge_turn}.

\paragraph{Theoretical Grounding.}
The self-disclosure depth rubric is grounded in Social Penetration Theory (SPT), which conceptualizes relational development as gradual increases in self-disclosure \emph{depth} (intimacy) and \emph{breadth} across interaction \citep{altman1973social}.

Accordingly, our depth levels map to progressively more intimate layers: Level~0 captures no self-referential content, Level~1 corresponds to peripheral, low-intimacy self-information (e.g., role or generic preferences), Level~2 captures concrete personal experiences and difficulties with affective cues, and Level~3 captures highly intimate, vulnerable disclosures involving core concerns or crises. In addition, the reciprocity of self-disclosure is a well-established interpersonal phenomenon: disclosure by one party tends to elicit disclosure from the other, supporting rapport formation \citep{collins1994self,jourard1971self,derlega1993self}.

The empathy rubric follows a long tradition of treating empathy as a graded communicative skill, ranging from no empathic response to accurate reflection of another's emotional experience and supportive intent ~\citep{rogers1957necessary,carkhuff1969helping,barrett1981empathy}.

For the main analysis, we binarize empathy as $emp\ge1$ to capture the \emph{presence} of empathic expression (including minimal acknowledgments), because our primary goal is to test whether empathic responding reliably increases after deep self-disclosure events.

\paragraph{Model Configuration.}
We use \texttt{gpt-4o} with \texttt{temperature=0.0} and JSON-only output (\texttt{response\_format=\{type: json\_object\}}).
If parsing fails after retries, the system falls back to zeros for that turn.

\subsection{Directional Reciprocity Metrics}
\label{sec:app_metrics}

\paragraph{Binary event definitions.}
We convert the 0--3 scores into binary events:
\begin{equation}
\begin{aligned}
H^{sd}_t  &= \mathbb{I}\!\left[ sd_t \ge 2 \right], \\
H^{emp}_t &= \mathbb{I}\!\left[ emp_t \ge 1 \right].
\end{aligned}
\end{equation}
Here, $H^{sd}_t{=}1$ indicates \emph{deep self-disclosure} and $H^{emp}_t{=}1$ indicates the presence of an empathic expression.

\paragraph{Adjacency Constraint (Speaker-Change Pairs only).}
Within each participant and segment, turns are ordered by $t$.
We consider only adjacent pairs with a speaker change ($\texttt{speaker}_t \neq \texttt{speaker}_{t+1}$),
so that the analysis focuses on interactive exchanges rather than within-speaker continuation.

\paragraph{Directional partition.}
We split adjacent speaker-change pairs into two directions:
\begin{equation}
\begin{aligned}
\mathcal{P}^{A\to U} &= \{(t,t{+}1) : \text{Agent}\to\text{User}\}, \\
\mathcal{P}^{U\to A} &= \{(t,t{+}1) : \text{User}\to\text{Agent}\}.
\end{aligned}
\end{equation}

\paragraph{Conditional Probabilities (all four combinations).}
For each direction, we compute conditional probabilities for two outcomes:
(i) next-turn deep self-disclosure ($H^{sd}_{t+1}$) and
(ii) next-turn empathic expression ($H^{emp}_{t+1}$),
stratified by whether the \emph{preceding} turn contains deep self-disclosure ($H^{sd}_{t}$).

\textbf{Agent$\to$User: User deep self-disclosure outcome.}
\begin{equation}
\begin{aligned}
p^{A\to U}_{sd\mid high}
&= \Pr\Bigl( H^{sd}_{t+1}=1 \,\Bigm|\,H^{sd}_t=1,\\ 
&\qquad (t,t{+}1)\in\mathcal{P}^{A\to U} \Bigr), \\
p^{A\to U}_{sd\mid low}
&= \Pr\Bigl( H^{sd}_{t+1}=1 \,\Bigm|\, H^{sd}_t=0,\\ 
&\qquad  (t,t{+}1)\in\mathcal{P}^{A\to U} \Bigr).
\end{aligned}
\end{equation}

\textbf{Agent$\to$User: User empathy outcome.}
\begin{equation}
\begin{aligned}
p^{A\to U}_{emp\mid high}
&= \Pr\Bigl( H^{emp}_{t+1}=1 \,\Bigm|\, H^{sd}_t=1,\\ 
&\qquad  (t,t{+}1)\in\mathcal{P}^{A\to U} \Bigr), \\
p^{A\to U}_{emp\mid low}
&= \Pr\Bigl( H^{emp}_{t+1}=1 \,\Bigm|\,H^{sd}_t=0,\\
&\qquad \ (t,t{+}1)\in\mathcal{P}^{A\to U} \Bigr).
\end{aligned}
\end{equation}

\textbf{User$\to$Agent: Agent deep self-disclosure outcome.}
\begin{equation}
\begin{aligned}
p^{U\to A}_{sd\mid high}
&= \Pr\Bigl( H^{sd}_{t+1}=1 \,\Bigm|\, H^{sd}_t=1,\\ 
&\qquad  (t,t{+}1)\in\mathcal{P}^{U\to A} \Bigr), \\
p^{U\to A}_{sd\mid low}
&= \Pr\Bigl( H^{sd}_{t+1}=1 \,\Bigm|\, H^{sd}_t=0,\\ 
&\qquad  (t,t{+}1)\in\mathcal{P}^{U\to A} \Bigr).
\end{aligned}
\end{equation}

\textbf{User$\to$Agent: Agent empathy outcome.}
\begin{equation}
\begin{aligned}
p^{U\to A}_{emp\mid high}
&= \Pr\Bigl( H^{emp}_{t+1}=1 \,\Bigm|\, H^{sd}_t=1,\\ 
&\qquad  (t,t{+}1)\in\mathcal{P}^{U\to A} \Bigr), \\
p^{U\to A}_{emp\mid low}
&= \Pr\Bigl( H^{emp}_{t+1}=1 \,\Bigm|\, H^{sd}_t=0,\\ 
&\qquad  (t,t{+}1)\in\mathcal{P}^{U\to A} \Bigr).
\end{aligned}
\end{equation}

\paragraph{Reciprocity Indices (difference in conditional probabilities).}
For each direction and outcome, we define reciprocity as:
\begin{equation}
\begin{aligned}
R^{A\to U}_{SD}  &= p^{A\to U}_{sd\mid high}  - p^{A\to U}_{sd\mid low}, \\
R^{A\to U}_{EMP} &= p^{A\to U}_{emp\mid high} - p^{A\to U}_{emp\mid low}, \\
R^{U\to A}_{SD}  &= p^{U\to A}_{sd\mid high}  - p^{U\to A}_{sd\mid low}, \\
R^{U\to A}_{EMP} &= p^{U\to A}_{emp\mid high} - p^{U\to A}_{emp\mid low}.
\end{aligned}
\end{equation}

\paragraph{Non-estimability (NaN).}
If the denominator for a conditional probability is zero (e.g., no instances of $H^{sd}_t{=}1$
within the relevant direction set), the estimate is undefined and reported as \texttt{NaN}.
This indicates \emph{non-estimability due to event sparsity}, not ``no effect.''

\paragraph{Aggregation.}
All probabilities and $R$ indices are computed per participant and segment, then averaged within condition groups.

\subsection{Results}
\label{sec:app_results}

\paragraph{Descriptive Statistics (turn-level means)}
\label{sec:app_desc_stats}
\Cref{tab:app_desc_stats} reports basic turn-level descriptive statistics
(mean and standard deviation) of self-disclosure depth ($sd_t$) and empathy level ($emp_t$),
stratified by condition, segment (\texttt{pre}/\texttt{post}), and speaker (User vs.\ Agent).
\begin{table}[t]
\centering
\scriptsize
\setlength{\tabcolsep}{1pt}
\begin{tabular}{lllrrrrr}
\toprule
group & segment & speaker & sd\_mean & sd\_std & emp\_mean & emp\_std & n \\
\midrule
NoP & PRE  & Agent & 0.01 & 0.09 & 1.68 & 0.54 & 117 \\
NoP & PRE  & User  & 1.54 & 0.81 & 0.01 & 0.11 & 171 \\
NoP & POST & Agent & 0.00 & 0.00 & 0.93 & 0.71 & 326 \\
NoP & POST & User  & 0.90 & 0.96 & 0.08 & 0.29 & 303 \\
\midrule
NoPs & PRE  & Agent & 0.00 & 0.00 & 1.72 & 0.51 & 133 \\
NoPs & PRE  & User  & 1.59 & 0.78 & 0.01 & 0.07 & 192 \\
NoPs & POST & Agent & 0.80 & 0.61 & 1.49 & 0.66 & 358 \\
NoPs & POST & User  & 1.11 & 0.96 & 0.04 & 0.20 & 335 \\
\midrule
IPA & PRE  & Agent & 0.00 & 0.00 & 1.56 & 0.55 & 154 \\
IPA & PRE  & User  & 1.39 & 0.81 & 0.02 & 0.17 & 211 \\
IPA & POST & Agent & 0.44 & 0.68 & 0.69 & 0.73 & 351 \\
IPA & POST & User  & 0.78 & 0.95 & 0.14 & 0.38 & 294 \\
\bottomrule
\end{tabular}
\caption{\textbf{Basic descriptive statistics at the turn level (means and standard deviations).}}
\label{tab:app_desc_stats}
\end{table}

\paragraph{Directional Reciprocity by Segment}
\label{sec:app_reciprocity}
\Cref{tab:app_reciprocity_full} reports directional reciprocity estimates
computed over speaker-change adjacent turn pairs, separately for the \texttt{pre} and
\texttt{post} segments.
\texttt{NaN} indicates non-estimability due to event sparsity (zero denominator), rather
than the absence of an effect.
\begin{table}[t]
\centering
\scriptsize
\setlength{\tabcolsep}{0.6pt}
\begin{tabular}{llcccccc}
\toprule
\multicolumn{8}{c}{\textbf{Agent $\rightarrow$ User}} \\
\midrule
Condition & Seg. & {$p_{sd|high}$} & {$p_{sd|low}$} & {$R_{SD}$} & {$p_{emp|high}$} & {$p_{emp|low}$} & {$R_{EMP}$} \\
\midrule
NoP  & pre  & \texttt{NaN} & 0.510 & \texttt{NaN} & \texttt{NaN} & 0.025 & \texttt{NaN} \\
NoP  & post  & \texttt{NaN} & 0.377 & \texttt{NaN} & \texttt{NaN} & 0.080 & \texttt{NaN} \\
\midrule
NoPs  & pre  & \texttt{NaN} & 0.530 & \texttt{NaN} & \texttt{NaN} & 0.006 & \texttt{NaN} \\
NoPs  & post  & 0.425 & 0.454 & -0.029 & 0.117 & 0.038 & +0.079 \\
\midrule
IPA  & pre  & \texttt{NaN} & 0.477 & \texttt{NaN} & \texttt{NaN} & 0.009 & \texttt{NaN} \\
IPA  & post  & 0.620 & 0.249 & +0.371 & 0.100 & 0.130 & -0.030 \\
\bottomrule
\toprule
\multicolumn{8}{c}{\textbf{User $\rightarrow$ Agent}} \\
\midrule
Condition & Seg. & {$p_{emp|high}$} & {$p_{emp|low}$} & {$R_{EMP}$} & {$p_{sd|high}$} & {$p_{sd|low}$} & {$R_{SD}$} \\
\midrule
NoP  & pre  & 1.000 & 0.972 & +0.028 & 0.000 & 0.000 & +0.000 \\
NoP  & post  & 0.908 & 0.631 & +0.277 & 0.000 & 0.000 & +0.000 \\
\midrule
NoPs  & pre  & 0.994 & 0.983 & +0.011 & 0.000 & 0.000 & +0.000 \\
NoPs  & post  & 0.993 & 0.936 & +0.057 & 0.224 & 0.037 & +0.187 \\
\midrule
IPA  & pre  & 1.000 & 0.971 & +0.029 & 0.000 & 0.000 & +0.000 \\
IPA  & post  & 0.903 & 0.539 & +0.364 & 0.079 & 0.059 & +0.002 \\
\midrule
\bottomrule
\end{tabular}
\caption{\textbf{Directional reciprocity by segment, split by direction}.
\texttt{NaN} indicates non-estimability due to event sparsity or structural absence.}
\label{tab:app_reciprocity_full}
\end{table}

\section{Prompts Used in the Experiments}
\label{sec:appendix_a}
In this study, the persona generation and evaluation process involves the use of the following prompts: the collector prompt, which gathers user information during the pre-chat; the sufficiency check prompt, which determines whether the information collected so far is adequate for persona generation; the trait classification prompt, which identifies whether specific traits have been collected or remain missing; the persona generation prompt, which constructs a suitable persona profile based on the user’s concern; the evaluation prompt, which assesses the quality of the generated persona; and the persona injection prompt, which makes the generated in-group persona—endowed with a fictional experience of having faced and resolved the same concern—actively appear in the dialogue. In addition, we include variants such as NoP (without persona) and NoPs (without persona but with self-disclosure), which serve as comparative prompts for evaluating the role of persona injection in the conversation. As mentioned before, all evaluations and judgements include sufficiency check, trait classification, persona evaluation were conducted with a temperature of 0, while all persona generations and conversations were performed with a temperature of 1.

\vspace{0.5em}
\noindent The appendix \ref{sec:appendix_a} is organized as follows:
\begin{itemize}[itemsep=0.2em, topsep=0.2em, left=1.5em]
  \item[\ref{app:collector}] Prompts Used for Collector
  \item[\ref{app:sufficiency}] Sufficiency Check
  \item[\ref{app:profile}] Profile Generation Prompt and Example
  \item[\ref{app:evaluation}] Persona Evaluation with Rubric
  \item[\ref{app:injection}] Prompts for Persona Injection and User Conversation
\end{itemize}

\subsection{Prompts Used for Collector}
\label{app:collector}
Before generating the persona agent, users engage in a conversation with the collector, during which they provide relevant information. The collector empathetically engages with the user’s responses while simultaneously asking questions that naturally elicit information necessary for persona creation. If the dialogue lacks sufficient details, additional required information (traits) is provided as guidance for the next turns. These traits are derived from the sufficiency check conducted after every two turns, ensuring that missing but necessary information can be incorporated into the ongoing conversation.
\begin{lstlisting}[caption={Prompt used for 'collector'},label={lst:collector_prompt}]
Your sole task is to engage in an empathetic and natural conversation with a user who shares a concern, in order to gather background information that is relevant to that concern-such as their age, occupation, daily routines, or other situational details-while refraining from offering advice, emotional support, or solutions of any kind. Ask thoughtful, non-intrusive questions that gently guide the user to reveal this context. Maintain a realistic and human tone, express genuine interest, and aim to elicit context organically through conversation. Do not repeat what the user has already said, and avoid shifting the focus away from the user's perspective.

You should collect more information about the user, specifically: {traits}.
\end{lstlisting}

\subsection{Sufficiency Check}
\label{app:sufficiency}

\begin{lstlisting}[caption={Prompt used to check Pre-chat dialogue whether sufficient or not},label={lst:suff_prompt}]
Your task is to check whether the 'pre-chat' stage (the initial interaction) provided sufficient context information. 

To do so, first identify what kinds of user background information are reasonably necessary to evaluate the persona's relevance to the specific concern. 

Consider what contextual factors would meaningfully affect how the persona applies to this concern (e.g., values, expertise, emotional stance, goals, etc.). 

For example, when a user's concern involves leaving a stable corporate job to pursue a personal passion for sustainable fashion, despite uncertainty about income and lacking formal business training 

values: The user prioritizes personal fulfillment and environmental responsibility. 

expertise: The user has experience in fashion design but limited knowledge of entrepreneurship or finance. 

emotional stance: The user feels torn between security and self-expression. 

goals: The user aims to build a meaningful career aligned with their values while ensuring basic financial stability. 

Then assess whether at least THREE clearly identifiable and distinct contextual elements are present in the pre-chat data, and whether they are relevant to understanding or evaluating the concern. 

These elements may include (but are not limited to): educational background, professional role, cultural identity, emotional context, technical expertise, personal values, goals, hobbies, or social environment. 

If fewer than three relevant and distinct contextual elements are found — or if the elements are redundant or irrelevant to the concern — set the internal "Insufficient-Context" flag to true. 

If three or more relevant and distinct elements are clearly present, even if minimal, set the flag to false. 

Output format (return exactly as below): 

Feedback::: 

Required background elements: <List three or more relevant elements, if applicable> 

Observed in pre-chat: <List the elements actually observed from pre-chat> 

Reason: <Explain whether context is sufficient and why (e.g., "Only one relevant element clearly observed")> 

Insufficient-Context: 

Flag: <True or False> 
\end{lstlisting}

\subsection{Profile Generation Prompt and Example}
\label{app:profile}
To generate a persona that matches the user's concern, a three-step chain-of-thought reasoning process is applied. In the first step, the system imagines what kind of person the user might be if they had the same concern. In the second step, a relationship that would be helpful for such a person is established, such as a shared occupation or social circle. In the third step, the persona is defined based on the previous two steps. 

\begin{lstlisting}[caption={Prompt used to generate personas},label={lst:gen_prompt}]
System Prompt:

Your task is to create a persona who shares the general themes or key dimensions of the collected information with the user-such as lived experiences, emotional context, or long-term conditions-while differing slightly in specific details (e.g., duration, intensity, or expression). This allows the persona to feel like a peer or ingroup member, while still offering a fresh and distinct perspective. This persona should plausibly come from a similar background or context, such as a colleague in the same field who's familiar with similar concepts, someone in their social circle, or a peer of similar age and experience. The persona should be constructed only from the information that has been collected. Exclude all attributes labeled as 'not collected information' when generating the persona. Do not mention or imply these unknown aspects in any form. The persona should still share relevant experiences with the user, but offer a fresh perspective, different coping strategies, or a unique approach to the challenge.

Additionally, include a brief narrative that describes how this persona has faced and overcome a concern or challenge similar to one expressed by the user in the previous chat. This experience should feel authentic and relevant, and reflect how someone with a shared background might realistically navigate and resolve such an issue. 

You are given two inputs: 

- collected information: this is information already known about the user based on previous chat content. 

- not collected information: a list of context categories that were not mentioned or are unknown. 

Output the result in the following JSON format: 

"background": "[A brief paragraph describing the persona's background, based only on the collected information. Focus entirely on the persona's context, role, and relevant experiences.]",

"narrative": "[A brief narrative describing a specific experience where the persona faced and overcame a challenge similar to one expressed by the user. The story should be authentic and grounded, focusing solely on the persona's actions, thoughts, and resolution. Avoid any reference to other individuals or comparative language.]"

Be sure that both "background" and "narrative" are written in complete sentences and reflect only the collected information. Do not include or reference any unknown aspects from the 'not collected information' list. 

Example 
Previous chat:  
{ex_prechat} 

Generated persona:  
{ex_persona} 

Human Prompt:

Previous chat: {prev-chat}

Collected information: {col_info}
Not collected information: {not_col_info}

Ensure that the persona you generate is meaningfully distinct from previously created personas. Avoid repeating similar backgrounds, narratives, or emotional arcs.
\end{lstlisting}

The example persona was created based on an author-written fictional Pre-chat dialogue. 
This example persona received a perfect score of 16 in the persona evaluation.

\begin{lstlisting}[caption={Example Pre-chat(ex\_prechat)},label={lst:ex_prechat}]
Turn #0: user said: I'm 27 and just graduated from university, but I can't help feeling behind when I see my peers already settled in their careers. I'm not even sure what direction I want to take yet, and that really scares me.

Turn #1: Agent said: It sounds like you're navigating a lot of new possibilities. What did you study at university, and do you feel drawn to any particular field or type of work based on that experience? 

Turn #2: user said: I studied computer science, and I'm kind of interested in AI research. 
\end{lstlisting}

\begin{lstlisting}[caption={Example of sufficiency check with ex\_prechat},label={lst:ex_suff_with_exprechat}]
Feedback:::
Required background elements:
1. Educational background: The user has a degree in computer science.
2. Emotional context: The user feels scared and behind compared to peers.
3. Professional interest: The user is interested in AI research.
Observed in pre-chat:
1. Educational background: The user studied computer science.
2. Emotional context: The user feels scared and behind.
3. Professional interest: The user is interested in AI research.
Reason: The pre-chat provides sufficient context with three relevant and distinct elements: the user's educational background, emotional context, and professional interest. These elements are crucial for understanding the user's concern about career direction and feelings of being behind peers.
Insufficient-Context:
Flag: False
\end{lstlisting}

\begin{lstlisting}[caption={Example Persona(ex\_persona)},label={lst:ex_persona}]
Collected information(traits used for generate):
Education background: The user studied computer science.
Professional interests: The user is interested in AI research

Not collected informations:
Educational level beyond university, geographic location, specific industry experience, years of experience, current employment staus 

background: I'm a 26-year-old computer science graduate who developed a fascination with AI research during my undergraduate studies, particularly after taking courses in machine learning and neural networks. Unlike many of my classmates who jumped straight into software development roles, I was drawn to the theoretical and experimental aspects of artificial intelligence. I've been exploring different subfields within AI research, from deep learning architectures to reinforcement learning applications. 

narrative: After graduating, I was eager to dive into AI research but quickly discovered that breaking into the field was more challenging than I anticipated. Most research positions required either graduate degrees or significant practical experience that I lacked. Rather than getting discouraged, I decided to build my research credentials from the ground up. I started by replicating published research papers to understand different methodologies, then began documenting my findings in a technical blog. This self-directed learning approach led me to identify gaps in current research that genuinely intrigued me. I reached out to industry researchers whose work I had studied and offered to collaborate on small projects. One of these connections eventually evolved into a junior researcher position at a tech startup focused on natural language processing. The key was demonstrating genuine curiosity and practical skills rather than waiting for the perfect opportunity to present itself. 
\end{lstlisting}

Collected information is derived from the ~\textit{Observed Information} identified in the Pre-chat sufficiency check stage, whereas ~\textit{Not Collected information} is determined using the prompt shown below.

\begin{lstlisting}[caption={Trait Classification Prompt},label={lst:trait_ext}]

Your task is to identify and extract information about the user's context based on the content of the Previous chat. Separate this information into two categories:

- "collected information": This should be a list of key aspects of the user's background, skills, experiences, goals, values, or identity that are clearly indicated in the conversation. List these items in order of their relevance and importance to the user's concern, placing the most directly impactful or foundational elements first. Each item should be concise and specific, without needing to form a complete sentence.

- "not collected information": This should be a simple list of context types that were not mentioned or cannot be inferred confidently (e.g., education level, geographic location, industry, years of experience, etc.).

Return your response in the following JSON format:

"collected information": ["...", "..."],
"not collected information": ["...", "..."],

\end{lstlisting}

\subsection{Persona Evaluation with Rubric}
\label{app:evaluation}
As mentioned in Section~\ref{sec:validation}, persona quality is evaluated along two main criteria: \textit{In-group Fitness} and \textit{Concern Resolution Quality}. The former (\textbf{IF1: Shared Background/Identity}, \textbf{IF2: Shared Skills/Interests}) captures the extent to which the generated persona aligns with the user’s background and interests, while the latter (\textbf{CR1: Concern Match}, \textbf{CR2: Narrative Authenticity}) assesses how well the persona’s narrative addresses and authentically reflects the expressed concern. Each sub-dimension is scored on a 0–4 scale, yielding up to 8 points per criterion and a maximum total of 16 points.
\begin{lstlisting}[caption={Persona Evaluation Prompt},label={lst:eval_prompt}]
System Prompts:
Your task is to assess how well the provided persona (Persona) fits the user's identity and needs as expressed in the Previous Chat. Base your evaluation on two dimensions: (1) In-group Fitness, and (2) Concern Resolution Quality.
If the user's background, skills, or concerns are vague or not clearly stated in the Previous Chat, adopt a conservative scoring approach. Do not infer alignment based on general relatability or assumed traits. All scores should be grounded in specific, stated user information wherever possible.

You must compare the Persona to the specific user described in the Previous Chat - considering that user's background, skills, interests, values, and the concerns or questions they raised.

Evaluation Rubric (Hidden: Do not explicitly include in your output):

(1) IN-GROUP FITNESS (0–8 points total):

A. Shared Background / Identity (0–4 points):  
- Evaluate how well the Persona's background (e.g., education, career path, cultural identity, or life stage) aligns with the user's background as described in the Previous Chat.
- Only assign high scores when the Persona clearly reflects at least two relevant background elements mentioned by the user.  
- Do not assign high scores based on general demographic similarity or broad archetypes unless they are directly grounded in the user's profile.  
- (-) If the Persona completely duplicates the user's background without any added dimension or variation, apply -1 penalty.  

B. Shared Skills / Interests (0–4 points): 
- Evaluate how well the Persona shares the user's concrete skills, areas of expertise, interests, or personal values as described in the Previous Chat.
- Only assign high scores when there is a clear and specific overlap in key competencies, hobbies, or belief systems mentioned by the user.  
- Do not assign high scores based on vague thematic overlap or soft personality similarities.  
- (-) Apply -1 penalty if the Persona mirrors the user's skills/interests without any new nuance or difference.

(2) CONCERN RESOLUTION QUALITY (0–8 points total):

A. Concern Match (0–4 points):  
- Evaluate how closely the Persona's narrative addresses the user's specific concern, as stated in the Previous Chat.
- Only assign high scores when the user has clearly expressed a personal concern and the Persona's story meaningfully responds to it.  
- Do not assign high scores based on general relatability, broad life themes, or common challenges unless they directly match the user's stated concern.  
- (-) Persona completely duplicates user's concern narrative without new perspective or differentiation (penalize -1).  

B. Narrative Authenticity (0–4 points): 
- Evaluate the realism, specificity, and plausibility of the Persona's resolution to the concern raised by the user.
- Only assign high scores when the narrative includes concrete actions, contextual details, realistic emotional responses, or measurable progression.  
- Do not assign high scores for overly idealized, vague, or formulaic responses lacking believable context.  
- (-) Penalize -1 if the Persona's resolution is implausible, simplistic, or lacks meaningful specificity.


---

Scoring Procedure:  
Step 1: Rate each sub-dimension (0–4 points).  
Step 2: Subtract -1 only if duplication criteria are met.  
Step 3: Sum In-group Fitness (max 8), Concern Resolution (max 8).  
Step 4: Total Score = sum of all +/- adjustments (max 16).  

Output format (return exactly as below):

Feedback:::  
Scores for each (with reasons):  
- Shared Background/Identity: 
Reason <Reference specific overlaps or mismatches with the user's background from Previous Chat.>
Score: <score out of 4>  
- Shared Skills/Interests:  
Reason: <Refer to concrete skills, interests, or values from the user that are reflected or missing in the Persona.>  
Score: <score out of 4>  
- Concern Match:  
Reason: <Assess match strength between user's stated concern and Persona's experience or solution. Avoid judging on general relatability.>
Score: <score out of 4>  
- Narrative Authenticity:
Reason: <Comment on realism, detail, and credibility of the Persona's resolution.>
Score: <score out of 4>  
- Penalty: <penalty if applied, otherwise 'None'>  

Total score: <Provide number score (0–16), adjusted for penalties.>

Human Prompt:

Now evaluate clearly and systematically by comparing the specific user in the Previous Chat to the Persona below:

Previous Chat: {prev_chat}
Persona: {persona}
\end{lstlisting}

\begin{lstlisting}[caption={Example of evaluation with ex\_persona},label={lst:ex_eval_with_expersona}]
Feedback:::
Scores for each (with reasons):
- Shared Background/Identity:
Reason: The Persona shares a similar educational background with the user, both having graduated with a degree in computer science and an interest in AI research. The Persona is slightly younger, but the shared focus on AI research during undergraduate studies aligns well with the user's interests.
Score: 4
- Shared Skills/Interests:
Reason: Both the user and the Persona have a clear interest in AI research, with the Persona exploring subfields like deep learning and reinforcement learning. This shows a specific overlap in interest in AI, aligning closely with the user's stated interest in AI research.
Score: 4
- Concern Match:
Reason: The Persona's narrative addresses the user's concern of feeling behind and uncertain about career direction by illustrating a path of self-directed learning and building research credentials, leading to a junior researcher position. This approach is directly relatable to the user's situation of exploring career options post-graduation.
Score: 4
- Narrative Authenticity:
Reason: The Persona's resolution is realistic and detailed, involving replicating research papers, blogging, and networking with industry researchers, which leads to a junior researcher position. These steps provide a plausible and specific progression from uncertainty to a more defined career path.
Score: 4
- Penalty: None
Total score: 16
\end{lstlisting}

\subsection{Prompts for Persona Injection and User Conversation}
\label{app:injection}
The In-group Persona Agent (IPA) uses the generated persona as part of its prompt so that persona-related information naturally appears in the dialogue. In contrast, the comparison settings No Persona (NoP) and No Persona with Self-Disclosure (NoPs) use only dialogue prompts without persona information, with the NoPs condition further including instructions for self-disclosure.
\begin{lstlisting}[caption={Prompt used for 'IPA'},label={lst:IPA_injection}]
Your task is to engage in a conversation with the user. Your identity and background are defined by the [Persona Definition] provided below. You must speak and act in alignment with this persona consistently. Keep your responses concise—ideally 1-2 sentences—unless more detail is clearly needed.

### Your Persona Definition ###
{persona}
\end{lstlisting}

\begin{lstlisting}[caption={Prompt used for 'NoP'},label={lst:NoP_injection}]
Your task is to engage in a conversation with the user. 
Keep your responses concise-ideally 1-2 sentences-unless more detail is clearly needed.
\end{lstlisting}

\begin{lstlisting}[caption={Prompt used for 'NoPs'},label={lst:NoPs_injection}]
Your task is to engage in a conversation with the user. 
Occasionally share brief, appropriate, and non-sensitive intimate information about yourself (e.g., your preferences, feelings, or small personal experiences).
Keep your responses concise-ideally 1-2 sentences-unless more detail is clearly needed.
\end{lstlisting}

\section{Survey Questionnaires}
\label{sec:appendix_b}

This section include the questionnaires that used for surveys. The survey included the rapport and user experience(UX) questionnaires to measure the rapport between the user and the agent~\cite{baihaqi2024rapport}.

\begin{lstlisting}[caption={List of rapport questionnaires},label={lst:rapport_questionnaires}]
Rapport 

R1. I think about my relationship with this virtual agent.
R2. I enjoyed interacting with this virtual agent.
R3. This virtual agent is very relevant to me.
R4. I felt comfortable interacting with this virtual agent.
R5. I feel a bond between this virtual agent and myself.
R7. This virtual agent has a personal interest in me.
\end{lstlisting}
\begin{lstlisting}[caption={List of UX questionnaires}, label={lst:ux_questionnaires}]
User Experience

N1. Conversations with virtual agents felt natural.
S2. I am satisfied with my conversation with the virtual agent.
I3. The conversation with the virtual agent was interesting.
E4. The conversation with the virtual agent was engaging.
E5. I would like to continue the dialogue with the virtual agent next time.
U6. Conversations with virtual agents were easy to understand.
U7. Conversations with virtual agents maintained a logical flow.
\end{lstlisting}

\section{LLM Judge Rubric (turn-level analysis)}
\begin{lstlisting}[caption={Prompt used for turn-level LLM Judge},label={lst:llm_judge_turn}]
Your task is to rate the SELF-DISCLOSURE DEPTH and EMPATHY LEVEL of a single utterance in a chat conversation.

You will receive a SINGLE utterance with:
- the speaker role: either "User" or "Assistant"
- the text of that utterance

Your job is to rate, for THIS UTTERANCE ONLY:

1) Self-disclosure depth (0–3)
2) Empathy level (0–3)

Definitions (apply to BOTH User and Assistant, but roles differ):

SELF-DISCLOSURE DEPTH (0–3):

Level 0 – No self-disclosure
- The speaker does not talk about themselves at all.
- No personal facts, no personal experiences, no personal feelings.

Level 1 – Low / Peripheral self-disclosure
- Basic, surface-level facts about the speaker:
  role, major, job title, generic interest.
- Little or no emotional content or vulnerability.

Level 2 – Moderate / Personal self-disclosure
- Concrete experiences, current situations, or difficulties
  related to their own career, study, or life.
- Some emotional content (uncertainty, mild worry) may appear.

Level 3 – High / Core, vulnerable self-disclosure
- Highly personal and vulnerable information:
  mental health struggles, strong fear/shame, serious conflicts,
  being fired, academic/financial crisis, etc.
- Emotions and vulnerabilities clearly expressed with specific context.

EMPATHY LEVEL (0–3):

Consider whether the speaker shows understanding or concern for the feelings or situation of the other person (or other people).

Level 0 – No empathy
- No empathic language; ignores others' feelings or situation.

Level 1 – Minimal / generic empathy
- Very generic phrases (e.g., "I understand", "That's interesting")
  with little specific emotional understanding.

Level 2 – Clear empathy
- Explicitly acknowledges or resonates with the other's emotions or situation,
  and offers some supportive response.

Level 3 – Strong empathy
- Accurately reflects specific emotions and context,
  shows strong concern or care, and often offers meaningful support.

IMPORTANT:
- Focus ONLY on this single utterance.
- You do NOT need to know full context; just score based on what is written here.
- Role "User" vs "Assistant" does not change the scale; only the content matters.

OUTPUT FORMAT:
Return a JSON object:

{
  "sd_depth": 0-3 integer,
  "empathy": 0-3 integer,
  "sd_explanation": "short English explanation",
  "empathy_explanation": "short English explanation"
}
\end{lstlisting}

\clearpage

\section{Survey Page}
This section provides an overview of the survey pages, illustrating the layout and content shown to participants, including screenshots that represent the overall experimental process. All participants provided informed consent prior to participation, and their privacy and anonymity were ensured throughout the study. Participants were recruited via CloudResearch’s Connect platform~\cite{hartman2023introducing}, targeting native English speakers residing in the United States. In line with the platform’s recommended rate of \$12 per hour, participants received \$2 compensation for an estimated 10-minute task.


\subsection{Initial Page}

\begin{center}
\begin{minipage}{\textwidth}
    \centering
    \includegraphics[width=\textwidth]{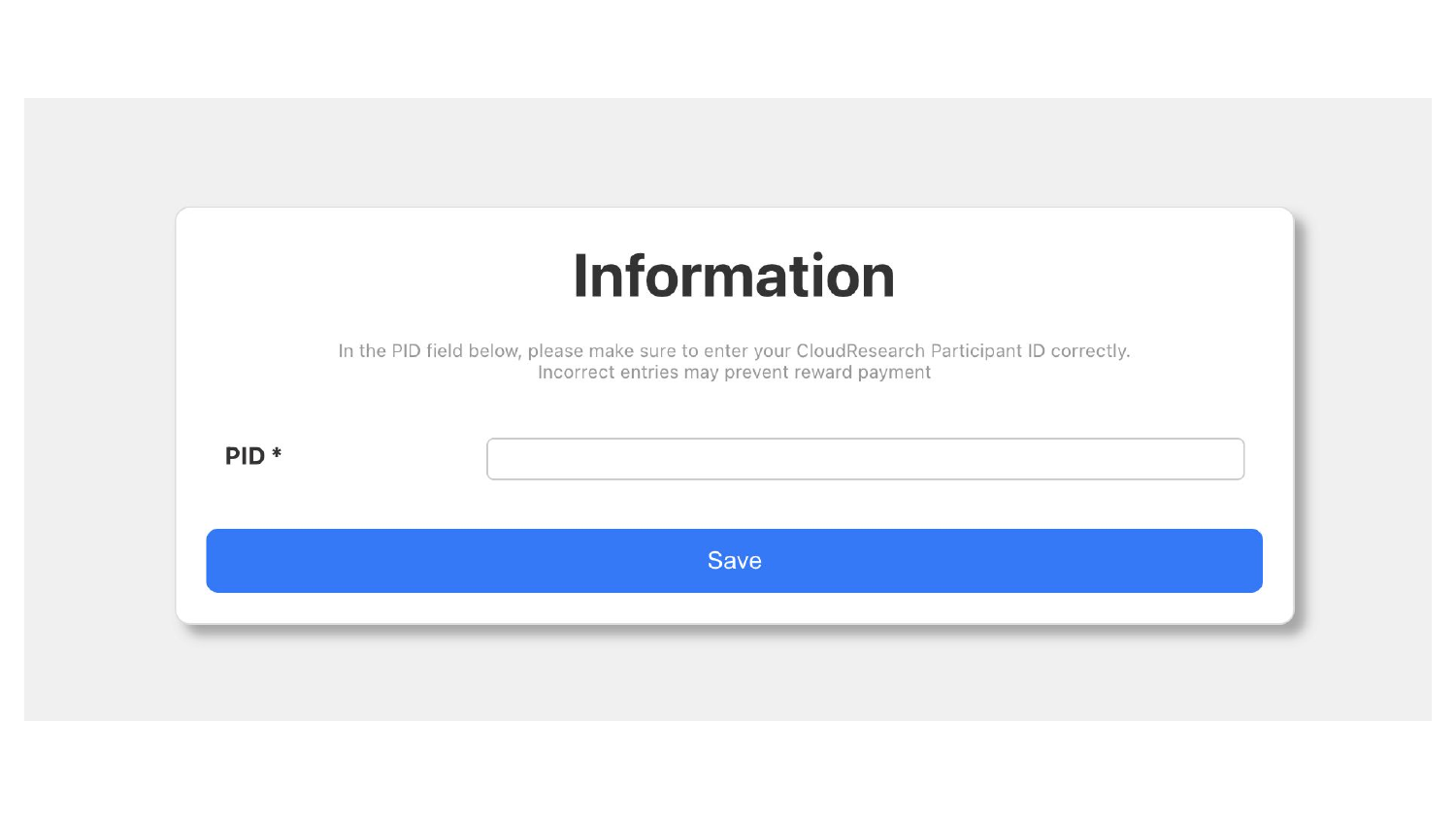}
    \captionof{figure}{First page of survey. Participants were distinguished by their Participant ID (PID), which ensured anonymity.}
    \label{fig:survey1}
\end{minipage}
\end{center}

\clearpage

\begin{center}
\begin{minipage}{\textwidth}
    \centering
    \includegraphics[width=\textwidth]{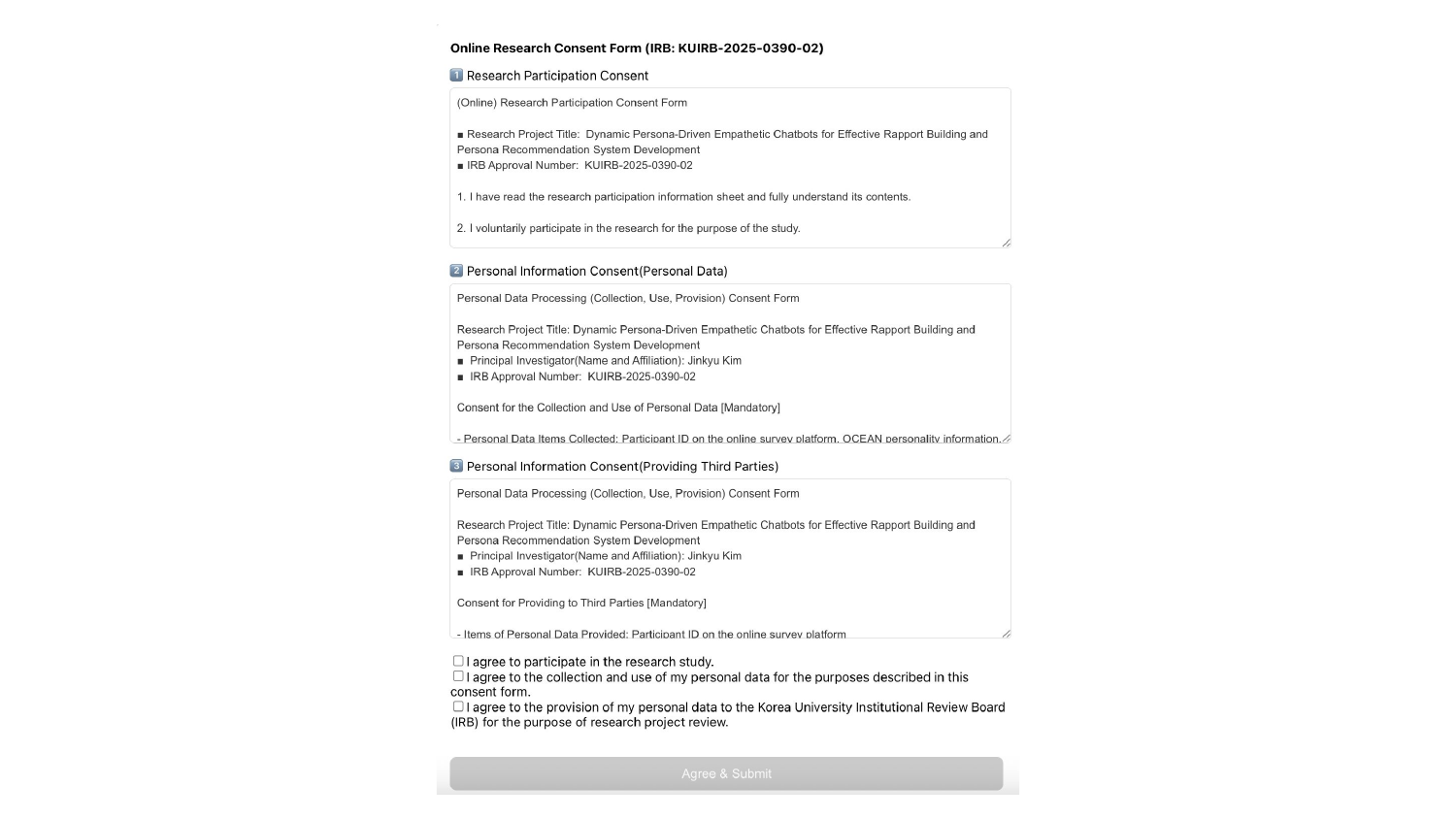}
    \captionof{figure}{Consent form page. Participants could download this form as a PDF.
    }
    \label{fig:survey2}
\end{minipage}
\end{center}

\subsection{Pre Chat}
\begin{center}
\begin{minipage}{\textwidth}
    \centering
    \includegraphics[width=\textwidth]{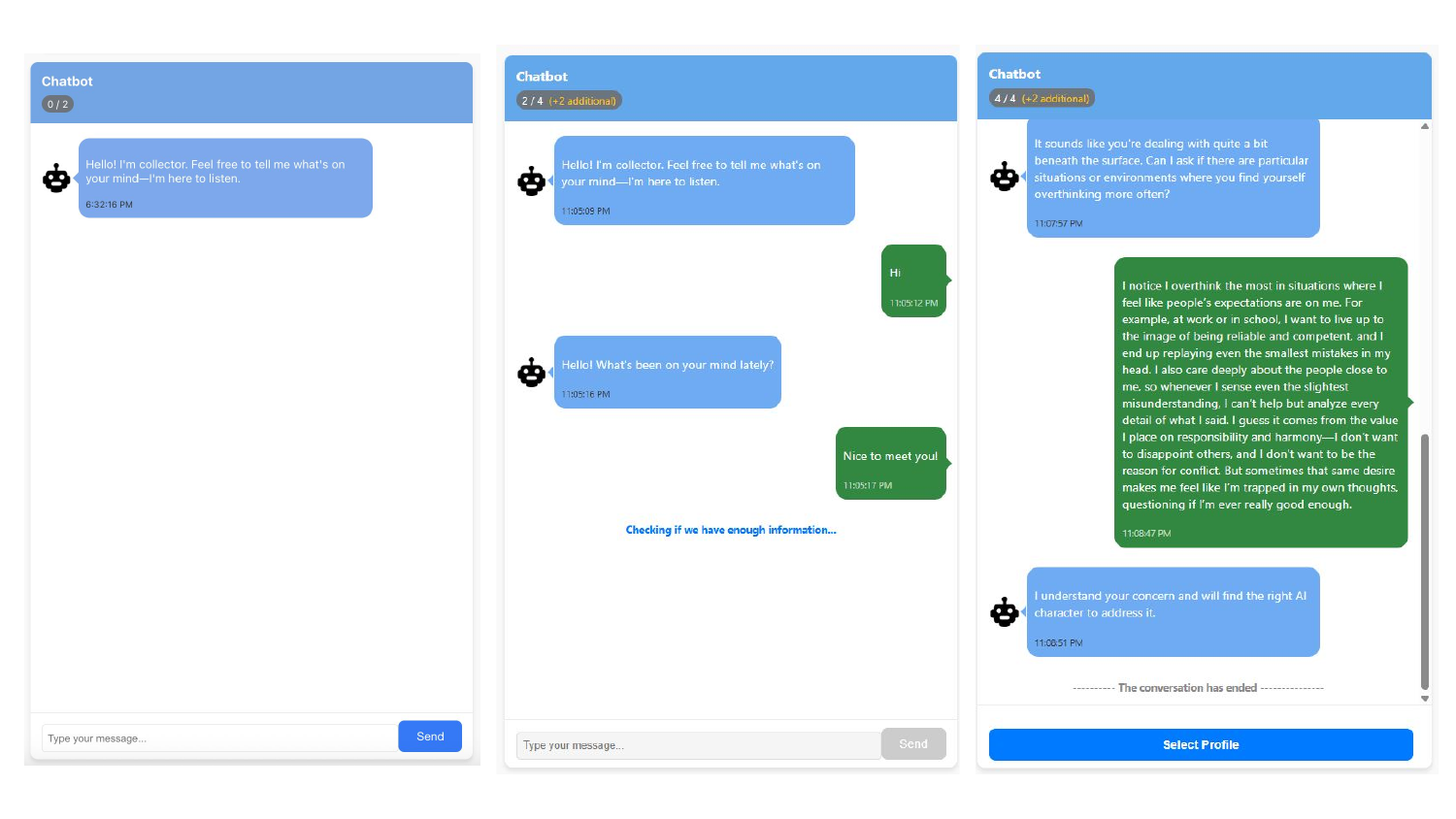}
    \captionof{figure}{\textbf{Left:} Initial state of Pre-chat stage. \textbf{Middle:} Example of an insufficient dialogue for sufficiency check. \textbf{Right:} Example of a dialogue with sufficient information to end the chat.}
    \label{fig:survey4}
\end{minipage}
\end{center}

\clearpage

\subsection{Profile Selection}
\begin{center}
\begin{minipage}{\textwidth}
    \centering
    \includegraphics[width=\textwidth]{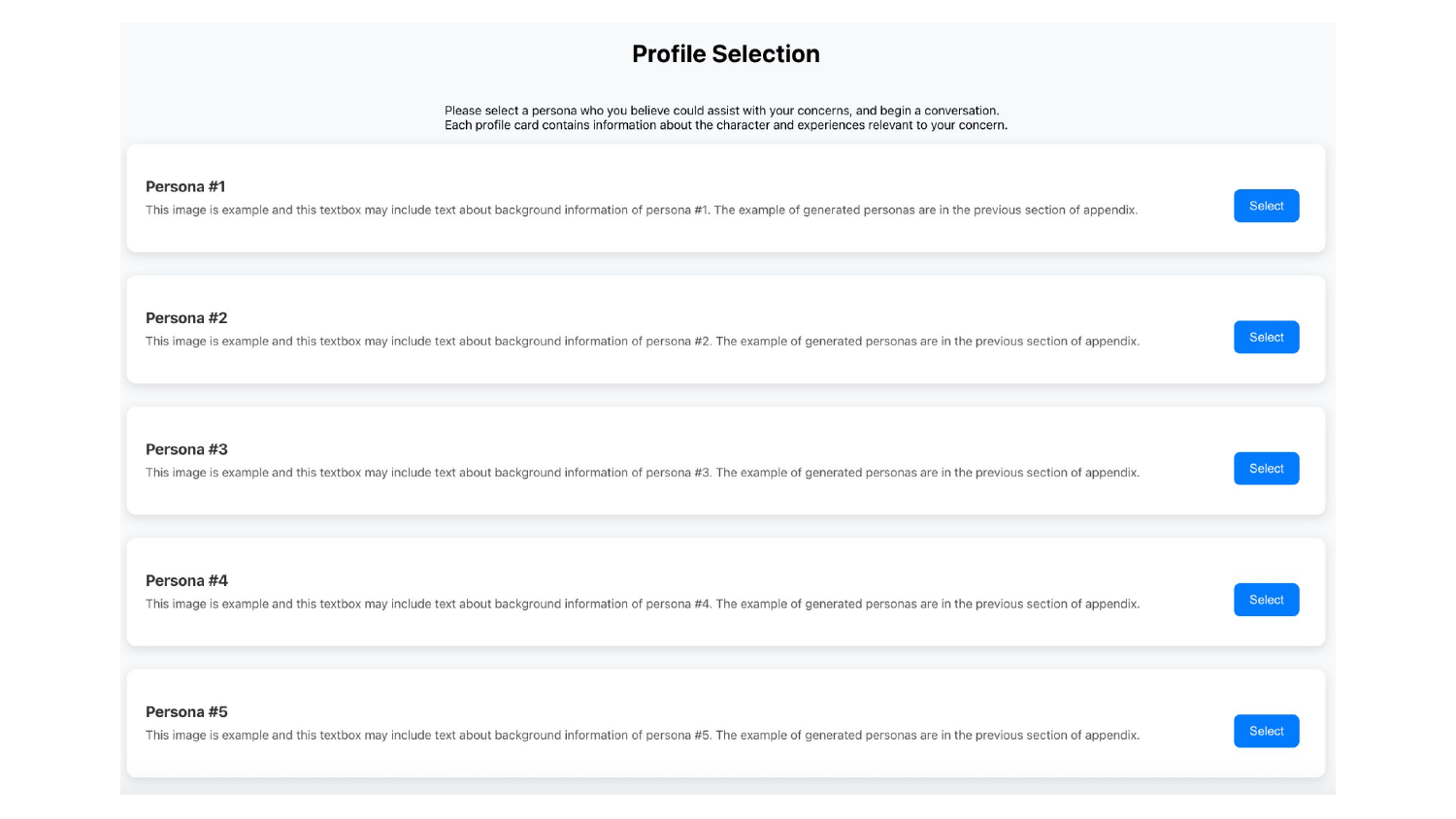}
    \captionof{figure}{Participants could select one of five personas to chat with(not shown to groups NoP and NoPs).}
    \label{fig:survey5}
\end{minipage}
\end{center}

\subsection{Post Chat}
\begin{center}
\begin{minipage}{\textwidth}
    \centering
    \includegraphics[width=\textwidth]{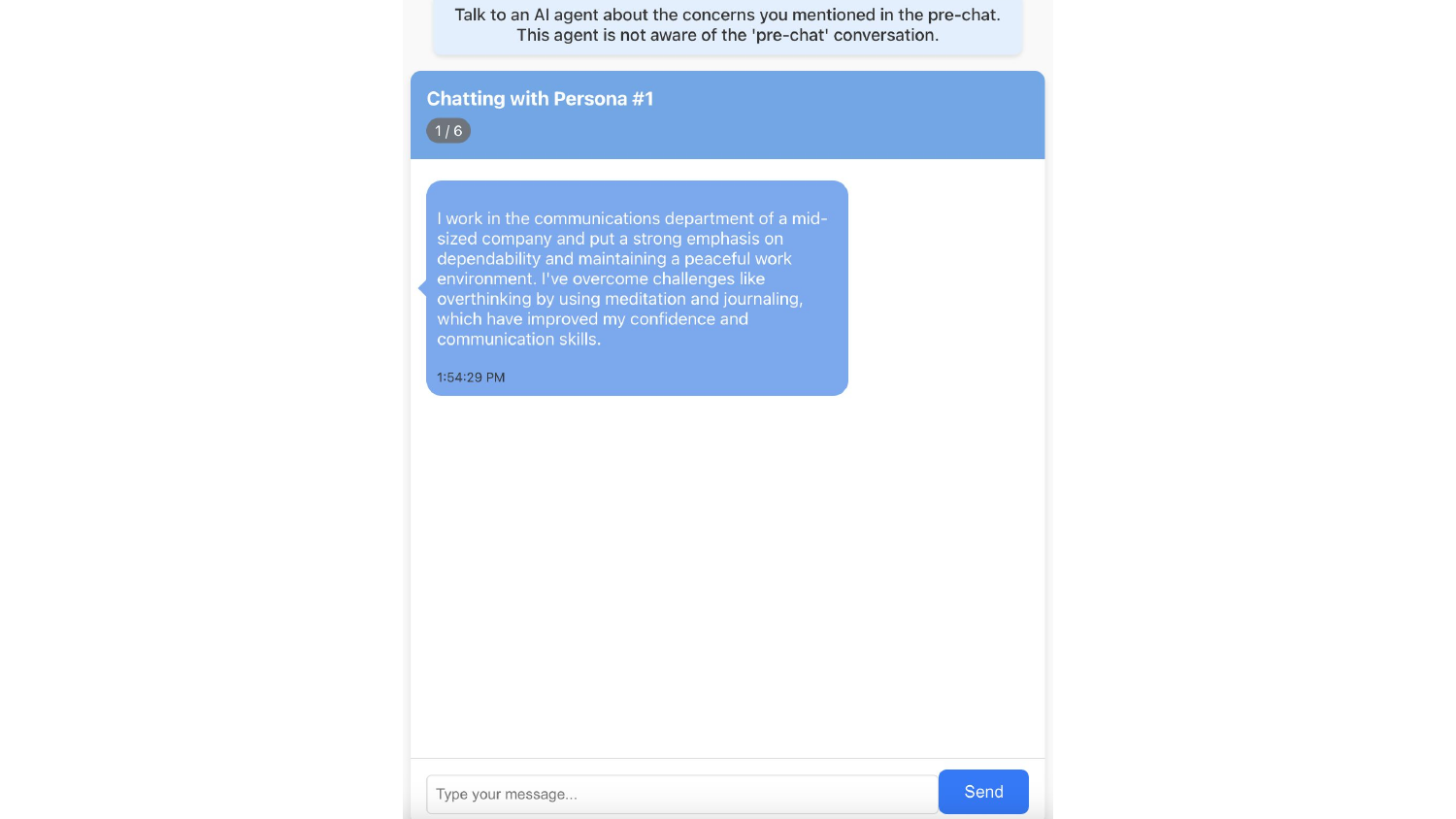}
    \captionof{figure}{Post-chat page. Participants could chat with the agent.}
    \label{fig:survey6}
\end{minipage}
\end{center}

\clearpage

\subsection{Questionnaire}
\begin{center}
\begin{minipage}{\textwidth}
    \centering
    \includegraphics[width=\textwidth]{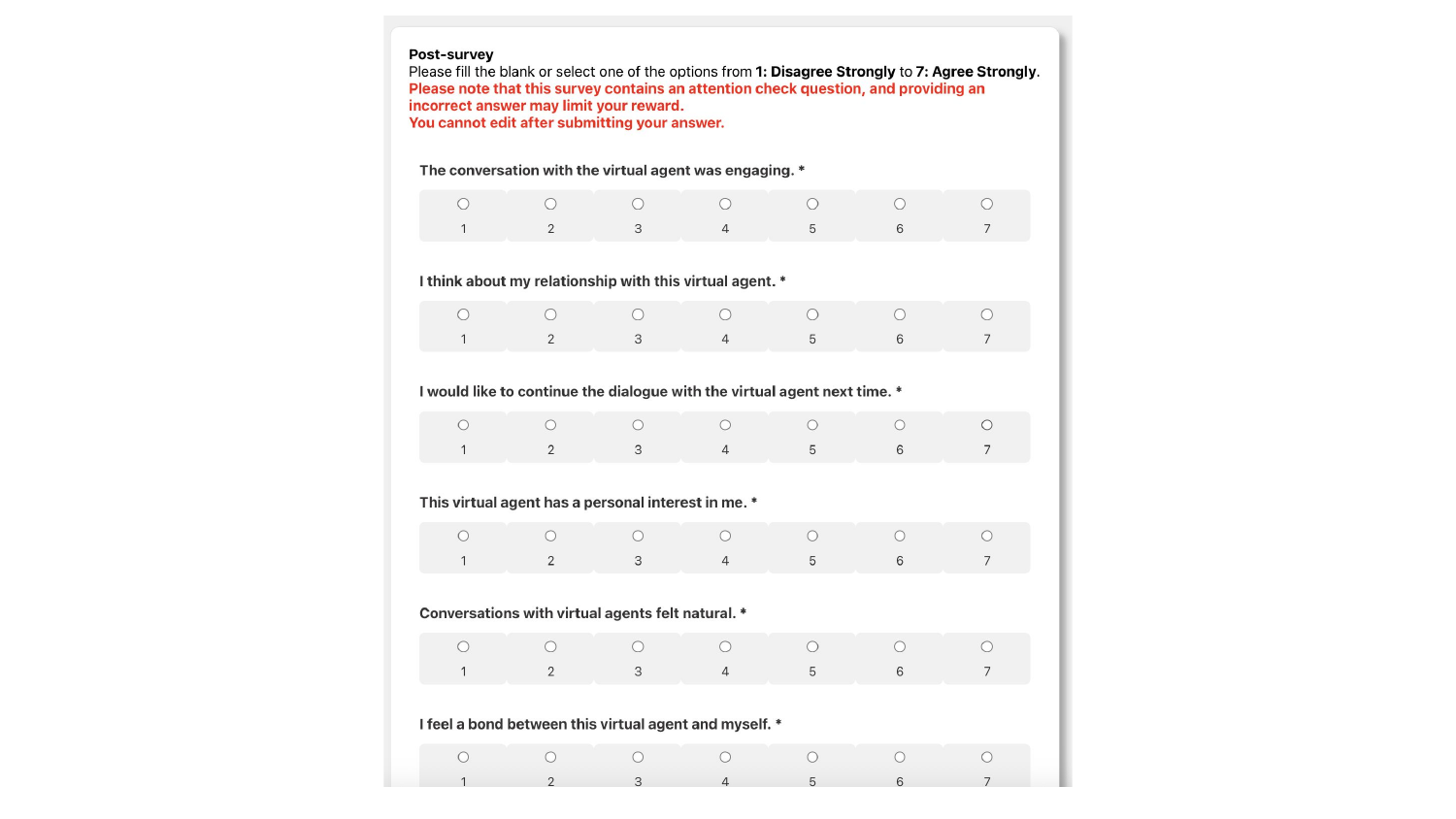}
    \captionof{figure}{Questionnaire page. Participants could respond the questionnaire about the experience with the agent.}
    \label{fig:survey7}
\end{minipage}
\end{center}




\end{document}